\documentclass[a4paper,fleqn,usenatbib]{mnras}

\usepackage{newtxtext,newtxmath}

\usepackage[T1]{fontenc}
\usepackage{ae,aecompl}

\usepackage{graphicx}		
\usepackage{amsmath}		
\usepackage{amssymb}		

\usepackage{xfrac}
\usepackage{cuted}		
\usepackage{anyfontsize}	

\title[Non-linear instability in eccentric discs]{Non-linear hydrodynamic instability and turbulence in eccentric astrophysical discs with vertical structure}

\author[A. F. Wienkers and G. I. Ogilvie]{
A. F. Wienkers$^{1,2}$\thanks{Contact e-mail: \href{mailto:wienkers@stanford.edu}{wienkers@stanford.edu}}\thanks{Present address: Flow Physics and Computational Engineering, Stanford University, Stanford, CA 94305} and
G. I. Ogilvie$^{2}$
\\
$^{1}$Cavendish Laboratory, Department of Physics, University of Cambridge, JJ Thomson Avenue, Cambridge CB3 0HE, UK \\
$^{2}$Department of Applied Mathematics and Theoretical Physics, University of Cambridge, Centre for Mathematical Sciences, \\ \quad Wilberforce Road, Cambridge
CB3 0WA, UK
}

\date{Accepted 2018 April 8. Received 2018 April 8; in original form 2017 November 16}

\pubyear{2018}

\begin{document}
\label{firstpage}
\pagerange{\pageref{firstpage}--\pageref{lastpage}}
\maketitle

\begin{abstract}
	
	
	Non-linear evolution of the parametric instability of inertial waves inherent to eccentric discs is studied by way of a new local numerical model.
	Mode coupling of tidal deformation with the disc eccentricity is known to produce exponentially growing eccentricities at certain mean-motion resonances. 
	However, the details of an efficient saturation mechanism balancing this growth still are not fully understood.
	This paper develops a local numerical model for an eccentric quasi-axisymmetric shearing box which generalises the often-used cartesian shearing box model.
	The numerical method is an overall second order well-balanced finite volume method which maintains the stratified and oscillatory steady-state solution by construction.
	This implementation is employed to study the non-linear outcome of the parametric instability in eccentric discs with vertical structure.
	Stratification is found to constrain the perturbation energy near the mid-plane and localise the effective region of inertial wave breaking that sources turbulence.
	A saturated marginally sonic turbulent state results from the non-linear breaking of inertial waves and is subsequently unstable to large-scale axisymmetric zonal flow structures. This resulting limit-cycle behaviour reduces access to the eccentric energy source and prevents substantial transport of angular momentum radially through the disc.
	Still, the saturation of this parametric instability of inertial waves is shown to damp eccentricity on a time-scale of a thousand orbital periods. It may thus be a promising mechanism for intermittently regaining balance with the exponential growth of eccentricity from the eccentric Lindblad resonances and may also help explain the occurrence of ``bursty'' dynamics such as the superhump phenomenon.
	
\end{abstract}

\begin{keywords}
	accretion, accretion discs -- hydrodynamics -- instabilities -- waves -- turbulence.
\end{keywords}

\defcitealias{OgilvieBarker:2014}{OB2014}
\defcitealias{BarkerOgilvie:2014}{BO2014}

\section{Introduction}

\subsection{Astrophysical motivation}

	
	
	The classical theory of accretion discs focuses on thin, circular, and coplanar viscous discs \citep{Pringle:1981gv}. 
	Yet there are no widely accepted local purely hydrodynamic instabilities in circular discs efficient enough to explain the deduced rate of mass accretion through observable discs.
	Thus by lifting the assumption of a circular disc, additional instability and transport processes may be excited.
	Indeed, although the lowest-energy orbit for a given angular momentum is circular, \citet{Syer:1992cc} have shown that not every disc tends to circularise.
	\citet{Lin:1976if} indeed have shown circularisation is inevitable for the special case of discs in a binary potential; however, in general this is dependent on the phase of the stresses in relation to the orbital phase.
	Analogous to the classic axisymmetric orbital diffusion problem for a circular ring, \citeauthor{Syer:1992cc} evolved an elliptical ring of matter, finding that the initial eccentricity can be sustained even on viscous time-scales for aligned orbits of similar eccentricities.
	Further mechanisms have more recently been found to couple with the evolution of eccentric disc modes and continually perturb or sustain eccentric orbits.
	These eccentric perturbations may be related to the origination of the disc, due to the continuous anisotropic mass transfer from a binary companion (Roche overflow) or from the parent nebular cloud core.
	Eccentricity may also be sustained by a viscous instability to eccentric perturbations which is known to exist for particular viscosity prescriptions, including the often-used $\alpha$-viscosity \citep{Lyubarskij:1994ep}.
	On longer time-scales, secular interactions with an eccentric companion or planet are also known to induce eccentricity.
	Finally, eccentric mode coupling with the tidal deformation via a mean-motion resonance has been shown to produce exponentially growing eccentricities when the disc and perturbing companion orbital frequency are nearly in a 
	ratio of $m$~to~$m\pm 2$ \citep{Lubow:1991eo,Lubow:2010fv}.
	These mean-motion resonances may arise in
		planetary rings resonating with satellites \citep{Goldreich:1981},
		protostellar discs with embedded planets \citep{Kley:2006er,Papaloizou:2001fb}, as well as
		circumstellar or circumbinary discs perturbed by a stellar binary \citep{Whitehurst:1988kn,Lubow:1991eo}.
	
	
	
	Thus the classical theory of accretion discs \citep{Pringle:1981gv} has been generalised to accommodate this additional degree of freedom in the orbital motion \citep{Ogilvie:2001ja}.
	Eccentric discs store additional energy in the $m=1$ azimuthal inertial mode, which opens up possibilities for instability and energy extraction not available to circular discs.
	One such instability arises from the parametric coupling between small-scale inertial waves and the periodically modulated parameters of the eccentric azimuthal mode, 
	generating growing inertial waves capable of transporting angular momentum \citep{Papaloizou:2005fl,BarkerOgilvie:2014}.
	The non-linear outcome of this instability is expected to be hydrodynamic turbulence resulting from the breaking of the inertial waves.
	Thus, this parametric instability may provide the necessary energy pathway to saturate the eccentricity growth, in particular from the strong $3$:$1$ eccentric inner Lindblad resonance viable in cataclysmic binaries, and which are shown to produce unmitigated eccentricity growth in linear theory \citep{Lubow:1991eo}.
	
	
	
	Large-scale zonal flows are found to arise out of this sustained turbulent state by way of non-linear mode coupling to large-scale perturbations which allows variation in Reynolds stresses over large length-scales.
	This is analogous to the driving of zonal flows out of magnetorotational turbulence \citep{Johansen:2009,Kunz:2013jp}.
	Both the magnetohydrodynamic and purely hydrodynamic driving mechanisms are critical agents in self-regulating their respective instabilities giving access to the free energy source. Thus these zonal flows have a significant impact on the transport quantities and global disc dynamics.
	Zonal flows also have a profound influence on the growth of planetesimals \citep{Dittrich:2013ix}.
	Relative flow to large solid bodies generates an aerodynamic drag which causes rapid drift through the disc, and can inhibit further accumulation and gravitational collapse into planetesimals \citep{Weidenschilling:1977hf}.
	However, \citet{Haghighipour:2003gn} found that the alternating super- and sub-Keplerian flows act as transport barriers, effectively trapping dusty matter and encouraging planetesimal growth \citep{Youdin:2002iza}.
	
	Overall, the details of this saturation mechanism may help explain the properties of superhumps and the evolution of superoutbursts in these binary systems, as well as the observed eccentricities of embedded planets \citep{Goldreich:2003}.
	There is also interest in finding a dependence of the $\alpha$-viscosity on disc eccentricity, which would contribute to higher-level analytic models allowing theorists to probe the early formation of protostellar discs, setting the initial conditions for star and planet formation.

\subsection{Computational background}

	These disc phenomena are able to be explored in a number of ways; however, upon exhausting linear perturbation theory, numerical models become an irreplaceable tool for studying the non-linear dynamics of these accretion flows.
	Current numerical methods for studying eccentric discs are limited to global simulations.
	Global smoothed particle hydrodynamics (SPH) simulations \citep{Whitehurst:1988kn,Murray:1995tq} were originally employed to deduce the global disc properties and eccentric instability modes; however, global finite volume and difference methods are now common \citep{Kley:2006er,Kley:2008,Marzari:2009dj}.
	Yet the parametric inertial wave instability has only ever been previously observed by \citet{Papaloizou:2005fo} using a global disc model with no vertical gravity. This is likely because the majority of global simulations of eccentric discs are either two-dimensional, and therefore unable to support this instability, or three-dimensional but too short in duration and numerically dissipative due to limits of computational resources \citep{Bitsch:2013}.
	Alternatively, simulating only a subset of the disc permits using higher spatial resolutions, as well as allowing more control over the numerical studies.
	Thus local models have seen many advances in theories of accretion discs where global simulations fall short. However, until now no generalisation of the shearing box has been implemented to model eccentric discs.

\subsection{Plan of this paper}

	%
	This paper is organised as follows.
	In Section \ref{sec: localEccentric} we review the local eccentric model of \citet{OgilvieBarker:2014} and develop derived quantities in terms of orbital averages needed to infer the global disc evolution. We then formulate the dispersion relation for inertial-acoustic waves and motivate a criterion for the breaking of inertial waves.
	In Section \ref{sec: numericalMethod} we modify the local model to develop a second order accurate $2.5$-dimensional finite volume scheme, and present a suite of validation tests.
	In Section \ref{sec: nonlinearSaturation} we present results of the numerical simulations and discuss the non-linear saturation of the parametric instability of inertial waves. 
	We further examine the ability for the saturated turbulence to damp the disc eccentricity and transport angular momentum.
	Finally in Section \ref{sec: largeScaleZonalFlows} we investigate the self-regulation of the parametric instability by large-scale zonal flows and formulate an activator-inhibitor dynamical model describing the interplay with the saturated turbulence.
	
	

\section{Eccentric local model}
\label{sec: localEccentric}

\subsection{Eccentric disc dynamics}
\label{sec: discDynamics}
	
	A global circular accretion disc model may be generalised to describe eccentric discs simply by changing the initial orbit equilibrium conditions.
	The initial orbital energy and angular momentum of a fluid parcel uniquely determine the eccentricity of the orbit. Consider a fluid parcel initially located at $r=\lambda$ with a specific orbital angular momentum $\ell = \sqrt{GM\lambda}$. Given a specific orbital energy equal to $\varepsilon_o = -GM/(2\lambda)$, the fluid parcel will take on a circular orbit; however, if the energy is increased to $\varepsilon_o = -\left(1-e^2\right)GM/(2\lambda)$, the orbit becomes eccentric with eccentricity $e$.
	Then to first order in $e$, the eccentricity appears as horizontal epicycles with frequency $\kappa$, which for a Keplerian disc, $\kappa = \Omega$.
	
	Thus a particular orbit is uniquely described by the semilatus rectum, $\lambda$, along with the orbit eccentricity.
	This suggests a generalisation of the polar coordinates, $(r, \phi)$, natural to eccentric discs, which was introduced by \citet{Ogilvie:2001ja}.
	These eccentric orbital coordinates, $(\lambda,\phi)$, therefore can describe any point in the disc when the eccentricity, $e(\lambda)$, and the longitude of pericentre, $\omega(\lambda)$, are also specified, and which are in general functions of $\lambda$.
	Because eccentric Keplerian orbits are ellipses, the transformation between polar and eccentric orbital coordinates is
	\begin{equation}
		r = \frac{\lambda}{1 + e \cos (\phi - \omega)}.
		\label{eq: transform}
	\end{equation}
	The Jacobian for this coordinate transformation from cartesian into the eccentric orbital coordinates is then
	\begin{equation}
		J(\lambda, \phi, z) = \frac{\partial (x,y,z)}{\partial (\lambda, \phi, z)} = \frac{\lambda \left( 1 + (e-\lambda e')\cos \theta - \lambda e \omega' \sin \theta \right)}{\left(1+e\cos\theta\right)^3}
		\label{eq: jacobian}
	\end{equation}
	where $e'$ and $\omega'$ are the derivatives with respect to $\lambda$ of the eccentricity and longitude of pericentre (or disc twistedness), respectively. $\theta = \phi - \omega$ here is the true anomaly, measuring the angular distance from pericentre.
	The Riemannian metric tensor, $g_{ij}$, and symmetric Levi-Civita connection, $\Gamma^i_{jk}$
	describe the vector calculus on this manifold, and were first derived for the orbital coordinates by \citet{Ogilvie:2001ja}. Each of these geometric coefficients for the eccentric orbital coordinates may be found in Appendix A of \citep[][hereafter OB2014]{OgilvieBarker:2014}.
	
	With the geometry now defined in the orbital coordinate system, the invariant form of the Euler equations in a gravitational potential can be written as
	\begin{subequations}
		\begin{align}
			D\rho &= -\frac{\rho}{J}\partial_i(Ju^i) \label{eq: invariantRho} \\
			Du^i + \Gamma^i_{jk} u^j u^k &= -g^{ij} \left( \partial_j \Phi + \frac{1}{\rho} \partial_j p \right) \label{eq: invariantU} \\
			Dp &= -\frac{\gamma p}{J} \partial_i(Ju^i), \label{eq: invariantP}
		\end{align}
		\label{eq: covariantHydro}
	\end{subequations}
	where $D \equiv \partial_t + u^i\partial_i$ is the Lagrangian derivative, $u^i$ is the typical contravariant velocity in the orbital coordinates, and $\Phi$ is the gravitational potential.
	An ideal gas equation of state, $p = \left(\gamma - 1\right) \rho \epsilon$, with adiabatic index, $\gamma$, was assumed in order to write equation (\ref{eq: invariantP}) in terms of pressure, providing closure. 
	Index notation with assumed summation over repeated indices is used to write (\ref{eq: covariantHydro}) succinctly.
	The additional connection terms in equation (\ref{eq: invariantU}) come from the covariant derivative in a general geometry, and will give rise to the Coriolis terms when entering a local frame.

\subsection{Local expansion}
\label{sec: localExpansion}
	
\begin{figure}
		\includegraphics[scale=1]{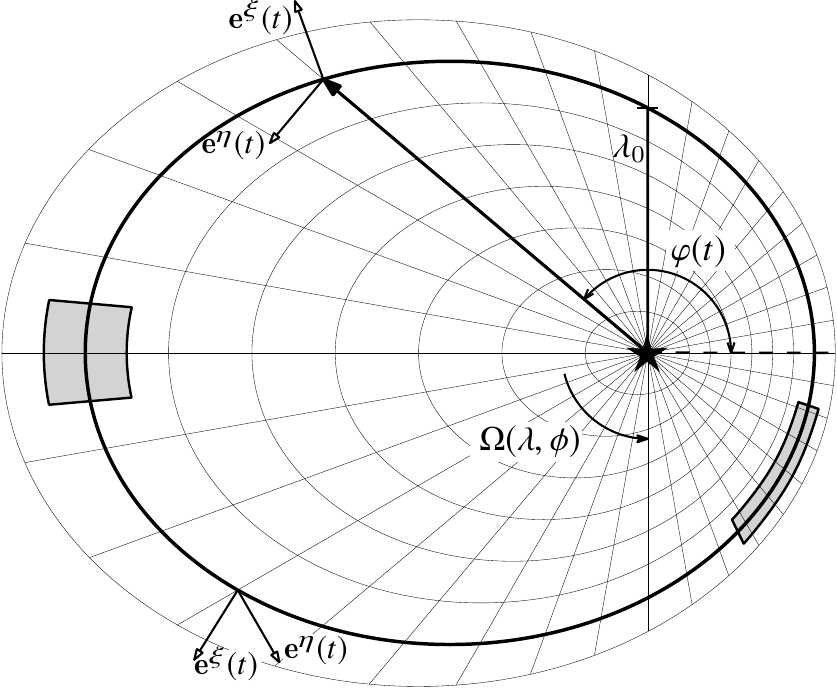}
		\caption[Global orbital coordinates on an eccentric disc.]{The global orbital coordinate grid, $(\lambda, \phi)$, shown overlaid on the disc. The fiducial orbit, $\lambda_0$, is shown in bold along with the local contravariant basis vectors, $(\mathbf{e}^\xi,\mathbf{e}^\eta)$, at two different times. The initial shearing box control volume at apocentre is shown stretched and advected at a later time.}
		\label{fig: eccentricDisc}
	\end{figure}

	Analogous to the local approximation used to formulate the cartesian shearing box model \citep{Goldreich:1965kv},
	the velocities in equation (\ref{eq: covariantHydro}) are decomposed into the background orbital motion, with $u^\phi = \Omega(\lambda,\phi)$, and the relative velocities, $v^i$, so that $(u^\lambda, u^\phi, u^z) = (v^\lambda, \Omega + v^\phi, v^z)$. 
	The global orbital coordinates are then expanded around a co-orbiting fiducial orbit, $\lambda_0$, centred on the mid-plane at $\varphi(t)$, and rotating with the disc at orbital frequency $\Omega(\lambda_0,\phi)$.
	These local non-orthogonal coordinates are $(\xi,\eta,z) = (\lambda - \lambda_0, \phi - \varphi(t), z)$, as shown at two different times in Fig.~\ref{fig: eccentricDisc}.
	
	\subsubsection{Governing equations}
	\label{sec: governingEquations}
	
		The local model developed in \citetalias{OgilvieBarker:2014} is further simplified by assuming local axisymmetry, thereby suppressing any non-zero global azimuthal modes. This follows from the assumption in the linear theory of \citet{BarkerOgilvie:2014} (hereafter BO2014) that the fastest growing azimuthal mode of the eccentric instability is the zero mode.
		Unfortunately, the generally non-orthogonal coordinate system means that a component of the $\partial_\xi p$ remains in the $\eta$-momentum equation, and so it cannot simply be passively advected as in the two-dimensional axisymmetric cartesian shearing box.
		The governing equations describing the \emph{local} quasi-axisymmetric model of an eccentric disc in primitive variables are then
		\begin{subequations}
			\begin{align}
				&D\rho = -\rho (\Delta + \partial_\xi v^\xi + \partial_z v^z) \label{eq: localEccentricRho} \\
				&Dv^\xi = -\frac{1}{\rho} g^{\lambda\lambda}\partial_\xi p - 2\Gamma^\lambda_{\lambda\phi}\Omega v^\xi - 2\Gamma^\lambda_{\phi\phi}\Omega v^\eta \label{eq: xiVelocity}\\
				&Dv^\eta = -\frac{1}{\rho} g^{\lambda\phi}\partial_\xi p - (\Omega_\lambda + 2\Gamma^\phi_{\lambda\phi}\Omega)v^\xi - (\Omega_\phi + 2\Gamma^\phi_{\phi\phi}\Omega)v^\eta \label{eq: etaVelocity}\\
				&Dv^z = -\Phi_2 z - \frac{1}{\rho} \partial_z p \\
				&Dp = -\gamma p ( \Delta + \partial_\xi v^\xi + \partial_z v^z),
			\end{align}
			\label{eq: localEccentric}
		\end{subequations}
		where the axisymmetric Lagrangian derivative is
		\begin{equation}
			D \equiv \partial_t + v^\xi \partial_\xi + v^z \partial_z,
		\end{equation}
		and the orbital velocity divergence is $\Delta \equiv J^{-1}\partial_\phi (J \Omega )$.
		
		Apart from the cross-over pressure gradient terms
			due to the in-plane coordinates being non-orthogonal,
		and the Coriolis terms
			coming from the connection terms of the covariant derivative with $u^\phi$,
		two additional forcing terms arise when considering the eccentricity.
		First, it is apparent that the effective tidal potential expansion now varies with $\phi$. Expanding the tidal potential similarly to the cylindrical case, but with $R(\theta) = \lambda_0 / (1+e\cos\theta)$, the second order term (in $z$) appearing in the local approximation is
		\begin{equation}
			\Phi_2 = \frac{GM}{\lambda_0^3}(1+e\cos \theta)^3,
			\label{eq: eccentricPhi}
		\end{equation}
		which is an implicit function of $t$, periodic with the eccentric orbital period,
		\begin{equation}
			\mathcal{P} = 2\pi \sqrt{\frac{\lambda_0^3}{GM}} \left( 1-e^2 \right)^{-\sfrac{3}{2}}.
			\label{eq: orbitalPeriod}
		\end{equation}
		This is a periodic variation in the vertical gravitational acceleration due to the time-varying radius while on a constant orbit, $\lambda_0$.
		
		A second modification to the cylindrical shearing box originates in the oscillatory geometry of the disc.
		Although the vertical ($\mathbf{\hat{z}}$) direction remains unchanged in the transformation, each of the velocity divergence terms in equations (\ref{eq: invariantRho}) and (\ref{eq: invariantP}) must be transformed into the new coordinate system, and are non-zero due to the azimuthal velocity of the fiducial orbit, $\mathrm{d}\varphi/\mathrm{d}t = \Omega$.
		Accounting for this horizontal velocity divergence amounts to including an additional energy and mass source in the governing equations.
		Using the Jacobian, $J(\lambda_0, \phi, 0)$, for the eccentric orbital coordinates, this velocity divergence on the $\lambda_0$ orbit can be expressed as
		\begin{equation}
			\Delta = \frac{1}{J} \frac{\partial}{\partial \phi} \left( J \Omega \right) = 
				\sqrt{\frac{GM}{\lambda_0^3}}\frac{\lambda_0(1+e\cos\theta)\left(e'\sin\theta - e \omega'(\cos\theta + e)\right)}{1+(e-\lambda_0 e')\cos\theta - \lambda_0 e \omega'\sin\theta}.
			\label{eq: divergence}
		\end{equation}
		This is again an implicit periodic function of time, with period $\mathcal{P}$,
		which arises from the misalignment of the apse lines in the disc ($\omega'$) in addition to the non-constant eccentricity ($e'$).
		The orbits illustrated in Fig.~\ref{fig: eccentricDisc} show this orbital convergence near pericentre, as well as the azimuthal elongation of the local domain between apocentre and just before pericentre due to the relative azimuthal acceleration of points on the same orbit.
		
		While the axisymmetric model presented above is quite general, as is that developed by \citetalias{OgilvieBarker:2014}, the following work will be limited to $e'=\omega'=0$ for isothermal discs ($\gamma = 1$).
		This greatly simplifies the following initial investigation into the non-linear evolution by reducing the number of disc control parameters from three down to only the eccentricity.

	\subsubsection{Horizontally invariant laminar solution}
	\label{sec: laminarSolution}
	
		The system of equations (\ref{eq: localEccentric}) admits laminar solutions that are horizontally invariant in the disc when $v^\xi = v^\eta = 0$. This system then reduces to a set of three equations which may be solved semi-analytically in the case of isothermal thermodynamics.
		
\begin{figure}
			\begin{center}
    			\includegraphics[scale=1]{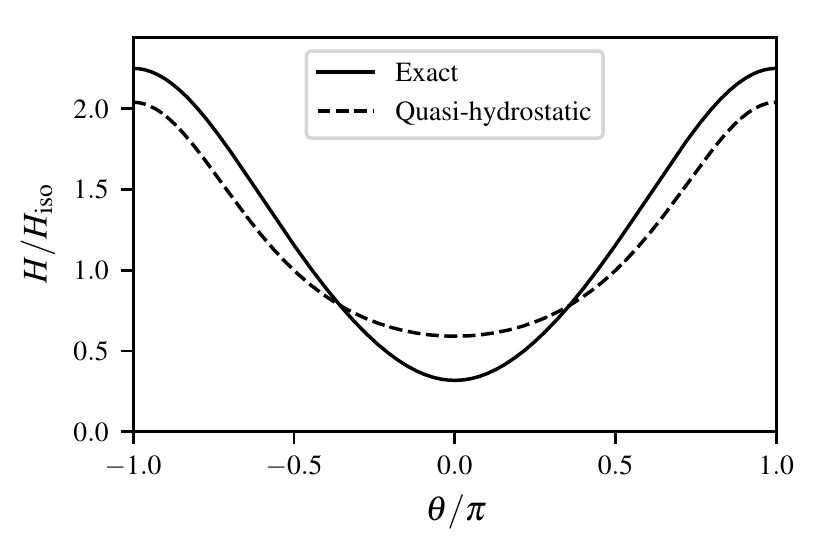}
    			\caption{Comparison of the resonant (laminar) solution for the thickness of the disc around an orbit 
							 with the na\"{\i}ve solution which uses the hydrostatic scale height at each orbital radius. Shown here for $e=0.3$.}
				\label{fig: quasiHydrostatic}
  			\end{center}
		\end{figure}

		A na\"{\i}ve quasi-hydrostatic approach to solving the resulting one-dimensional system might involve finding the isothermal scale height for a cylindrical disc at radius $R(\theta) = \lambda_0 / (1+e\cos\theta)$ as it varies about the orbit.
		Although incorrect due to neglecting all dynamics of the system, it is still instructive for demonstrating the dynamical amplification (shown in Fig.~\ref{fig: quasiHydrostatic} for a disc with $e=0.3$).
		
		A semi-analytic solution becomes possible after noticing that the eccentric forcing terms generate homogeneous vertical oscillations with period $\mathcal{P}$.
		The primitive variables are rewritten to uncouple the temporal and spatial dependencies by exploiting the solution form of the homogeneous oscillations,
		\begin{subequations}
			\begin{align}
				\rho(t,z) &= \hat{\rho}(t) \mathrm{e}^{-\zeta^2/2} \label{eq: rhoSimilarity}\\
				v^z(t,z) &= \hat{v}^z(t) z \\
				p(t,z) &= \hat{p}(t) \mathrm{e}^{-\zeta^2/2} \label{eq: pSimilarity}
			\end{align}
			\label{eq: primitivesForm}
		\end{subequations}
		where the similarity variable, $\zeta(t) \equiv z/H(t)$, can be defined for an isothermal disc.
		The governing equations are then uncoupled by inserting this form for the primitive variables into~(\ref{eq: localEccentric}):
		\begin{subequations}
			\begin{align}
				\frac{\dot{\hat{\rho}}}{\hat{\rho}} - \frac{H}{\dot{H}}\zeta^2 - \hat{v}^z\zeta^2 &= -(\Delta + \hat{v}_z) \\
				\dot{\hat{v}}^z + (\hat{v}^z)^2 &= -\Phi_2 + \frac{c_s^2}{H^2},
			\end{align}
			\label{eq: uncoupledODE}
		\end{subequations}
		where the redundant pressure equation is dropped.
		However, as the additional variable $H(t)$ has been introduced, a final relation is needed to close the system.
		This is found using a global mass conservation argument motivated by (\ref{eq: localEccentricRho}). Integrating this term vertically over the disc and using the form (\ref{eq: rhoSimilarity}), the closing equation is
		\begin{equation}
			\frac{\dot{\hat{\rho}}}{\hat{\rho}} + \frac{\dot{H}}{H} = -\Delta.
			\label{eq: globalMassConservation}
		\end{equation}
		This set can now be simplified in terms of $H$, giving a second order non-linear ODE describing the evolution of $H$:
		\begin{equation}
			\ddot{H} = -\Phi_2 H + \frac{c_s^2}{H}.
			\label{eq: odeH}
		\end{equation}
		This ODE describes a non-linear oscillator periodically forced by $\Phi_2$, 
		and admits both free oscillations as well as a forced response driven at the orbital frequency.
		The purely forced response can then be obtained by writing this as a periodic boundary value problem that can be solved numerically for $H$, and for which $\hat{\rho}$, $\hat{v}_z$, and $\hat{p}$ easily follow.
		
		Thus in a local model, an eccentric disc exhibits a vertical oscillatory ``breathing'' mode, where the disc scale height periodically expands and contracts as shown in Fig.~\ref{fig: quasiHydrostatic}.
		It is this vertical oscillation mode which, in addition to the oscillating horizontal orbital geometry, parametrically couples with inertial waves and causes resonant instability.
		Nonetheless, these horizontally invariant laminar solutions appear stationary in the inertial frame and therefore describe the global steady state of the eccentric disc, which will prove necessary when constructing and validating the numerical model.

	\subsubsection{Turbulent transport}
	\label{sec: turbulentTransport}
	
		It is now of interest to deduce the mass and angular momentum transport rates for the global disc represented by this local model.
		The angular momentum transport is often parameterised by the $\alpha$-viscosity parameter \citep{Shakura:1973uy}, which posits that the efficiency with which angular momentum can be transported through the disc is proportional to the local averaged pressure.
		Thus $\alpha$ non-dimensionalises the in-plane shear stresses responsible for transport with the average pressure, and for Keplerian shear is
		\begin{equation}
			\alpha =  -\frac{2}{3} \frac{T^{r\phi}}{p}.
			\label{eq: alphaCircular}
		\end{equation}
		
		As explicit viscous stresses are not considered here and the flow is assumed to be locally axisymmetric, the specific angular momentum of each fluid parcel is conserved around an orbit.
		Thus the only source of $\alpha$ 
			can be seen as being attributed to some enhanced turbulent viscosity acting on the background shear,
		\begin{equation}
			 \nu_T \sim \frac{\alpha c_s^2}{\Omega}.
		\end{equation}
		In steady turbulence this mixing is given by the Reynolds stresses,
		\begin{equation}
			T_R^{ij} = -\left< \rho v^i v^j \right>,
		\end{equation}
		where $\left<\cdot\right>$ indicates a time-average (or equivalently an azimuthal average).
		This turbulent stress arises from correlated turbulent velocity fluctuations, and often greatly exceeds viscous stresses in developed turbulence.
		Ignoring viscous stresses, the total stress can then be written for a general geometry as
		\begin{equation}
			T^{ij} = -\left< \rho v^i v^j \right> - g^{ij}p,
		\end{equation}
		where it should be noted that for a general non-cartesian metric (e.g. in orbital coordinates), there is an off-diagonal pressure contribution to the shear stress.
		
		In an eccentric disc, the in-plane shear stress, $R T^{\lambda \phi}$, is the component responsible for turbulent transport of angular momentum ``radially'' through the disc.
		More specifically, the consequent internal torque, $R^2 T^{\lambda \phi}$, rather must be used to generalise (\ref{eq: alphaCircular}) for an eccentric disc by properly accounting for variations in the radius around the orbit.
		The net effect may then be found by integrating the torque and total pressure over the entire disc,
			giving the generalisation of the $\alpha$-viscosity,
		\begin{equation}
			\alpha = - \frac{ \iint J R^2 T^{\lambda \phi}  \, \mathrm{d}\varphi \, \mathrm{d}z  }{  \iint J \lambda_0 p \, \mathrm{d}\varphi \, \mathrm{d}z }.
			\label{eq: alphaViscosity}
		\end{equation}
		This global azimuthal integral is calculated in the local model by substituting $\mathrm{d}\varphi = \Omega \, \mathrm{d}t$ and integrating over a single orbital period.
		It should be noted that this generalisation of the $\alpha$-viscosity still reduces to the familiar form for a circular disc, as $R = \lambda_0$ drops out of the integral, and thus requires no distinction between torque and in-plane shear stress.

	\subsubsection{Eccentricity evolution}
	
		Once a sustained turbulent state is achieved in the local model, it is necessary to determine the induced eccentricity decay rate balancing the eccentricity growth from mean-motion resonances. 
		This is not as simple as equating the dissipation of turbulent energy with an orbital energy decrease, consequently producing eccentricity decay.
		Depending on the phase of stresses in the orbit, the dissipation of perturbation energy may either decrease the angular momentum of the disc (thereby increasing eccentricity), or reduce the orbital energy (decreasing eccentricity). 
			
		Equation (52) in \citetalias{OgilvieBarker:2014} describes the precession and evolution of the global disc eccentricity, but requires the mean quasi-radial velocity through the disc in order to solve.
		Assuming a uniformly eccentric ($e'=\omega'=0$) and non-precessing disc, the eccentricity decay rate may be simplified to
		\begin{equation}
			\begin{aligned}
			\ell_0 \mathcal{M} \partial_t e =
						\iint \biggl[  2\left( \cos\theta + e \right) \partial_\lambda \left( J R^2 T^{\lambda\phi} \right)
										 - \frac{J R^2}{\lambda}\left( \cos\theta + e \right) T^{\lambda\phi}& \\
									 +\, \lambda \sin\theta\, \partial_\lambda\left( J R_\lambda T^{\lambda\lambda}\right)
										+ J R^2 \sin\theta\, T^{\phi\phi}  \biggr] \, \mathrm{d}\theta \, \mathrm{d}z, &
				\label{eq: eccentricityEvolutionUnsimplified}
			\end{aligned}
		\end{equation}
		where $\ell_0$ 
		is the specific angular momentum of the fiducial orbit, and $\mathcal{M} = \iint J\rho\Omega \, \mathrm{d}t \, \mathrm{d}z$ is the total mass of the disc.
		In order to evaluate the $\partial_\lambda$ terms, additional approximations about the radial disc structure must be made.
		A global steady alpha-disc model with constant $\alpha$ is often used to describe accretion discs, and is likely a good approximation for discs in cataclysmic binaries \citep{Shakura:1973uy}.
		This model produces a flared disc with surface density scaling $\Sigma \propto r^{-\sfrac{3}{4}}$, and vertically-integrated pressure $P \propto r^{-\sfrac{3}{2}}$.
		This same model was assumed in the work by \citet{Lubow:2010fv} to find the eccentricity growth rate in tidally distorted discs, and so will also be used here to facilitate comparison to the eccentricity decay rate.
		Another simple self-similar model for an accretion disc with a constant aspect ratio, $\Sigma \propto r^{-\sfrac{5}{2}}$, and the same scaling for $P$ as the alpha disc may also be a reasonable model. Regardless, the sum in (\ref{eq: eccentricityEvolutionUnsimplified}) is found to be only weakly dependent on this choice.
		
		If the stress (with respect to a normalised basis) scales with $\lambda$ as $\lambda^{-\sfrac{3}{2}}$, as suggested by a steady alpha disc, then the $T^{\lambda \phi}$ terms cancel exactly in (\ref{eq: eccentricityEvolutionUnsimplified}), giving
		\begin{equation}
			\ell_0 \mathcal{M} \partial_t e =
						\iint J\sin\theta \left( -\frac{1}{2} R_\lambda T^{\lambda\lambda} + R^2 T^{\phi\phi} \right) \Omega \, \mathrm{d}t \, \mathrm{d}z.
			\label{eq: eccentricityEvolution}
		\end{equation}
		Thus for this steady alpha-disc model, 
		the eccentricity evolution is dictated solely by the $\xi$ and $\eta$ normal stresses varying around the disc.
		For this reason, it is expected that the $\alpha$-viscosity and eccentricity decay rates are greatly influenced by the nature of the sustained turbulence. This steady turbulent state sourced by the non-linear breaking of inertial waves must now be considered.

\subsection{Inertial-acoustic waves}
\label{sec: inertialAcousticWaves}

\begin{figure}
		\begin{center}
    		\includegraphics[scale=1]{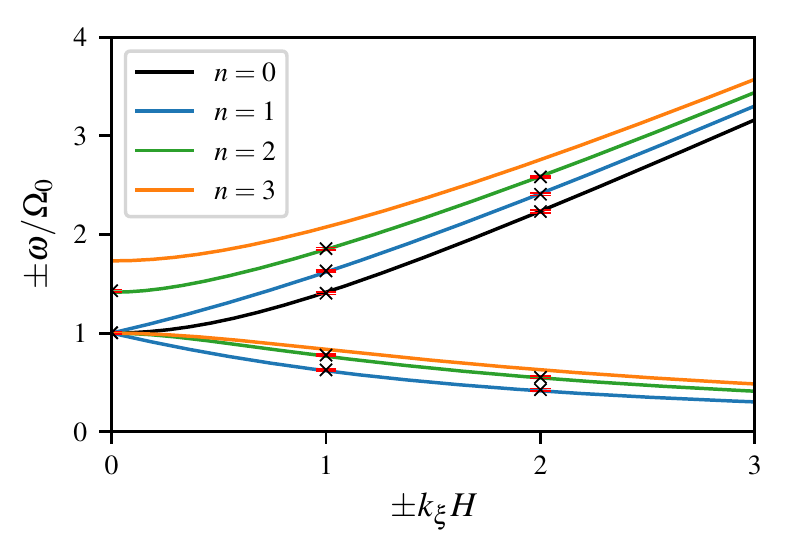}
    		\caption{Inertial-acoustic dispersion relationship (\ref{eq: dispersion}) giving the viable freely-propagating waves,
						$\omega(k_\xi)$, for the first four modes in a circular disc. x's indicate the numerically supported waves during validation runs. }
			\label{fig: dispersion}
  		\end{center}
	\end{figure}

	Internal waves are propagating disturbances whose restoring force may be gravity or Coriolis, and which are permitted to propagate within a fluid by some stable density stratification or rotation.
	In a rotating frame, and where the dynamical time-scale is comparable to the orbital frequency, inertial-gravity waves are supported by the influence of the Coriolis force \citep{Greenspan:1968uu}.
	These inertial (``epicyclic'') waves freely propagate only at particular frequencies given by the dispersion relation which depends on the spatial length-scale of the perturbation.
	A linear perturbation analysis of the governing equations in the local coordinates with axisymmetric perturbations of the form
	$ \hat{\delta}(z)\exp \mathrm{i} ( k_\xi \xi - \omega t) $
	results in the dispersion relation for a circular disc \citep{Okazaki:1987}:
	\begin{equation}
		(\omega^2 - n\Omega^2)(\omega^2 - \kappa^2) = \omega^2 k_\xi^2 c_s^2,
		\label{eq: dispersion}
	\end{equation}
	where $n \in \mathbb{Z}^+$ designates the vertical mode.
	Fig.~\ref{fig: dispersion} shows the viable solutions for the first four modes of the dispersion relation.
	Two distinct branches are evident for these inertial-acoustic waves in the limit of large $k_\xi$:
		The high frequency branch has an acoustic character in which $\omega/\Omega_0 \to k_\xi H$.
	In other words, these waves are non-dispersive and the phase speed $v_p = c_s$ similar to acoustic waves.
		In contrast, the low frequency branch appears as an inertial wave, which for large wavenumber gives $\omega/\Omega_0 \to \sqrt{n}/(k_\xi H)$.
	
	Finally, the vertical structure of the eigenmodes $\hat{\delta}(z)$ can be found in terms of Hermite polynomials \citep{Okazaki:1987,2013MNRAS.433.2420O}:
	\begin{subequations}
		\begin{align}
			\hat{v}^\xi, \hat{v}^\eta, \hat{h} &\propto \mathrm{He}_n\left(\frac{z}{H}\right) \\
			\hat{v}^z &\propto \mathrm{He}_{n-1}\left(\frac{z}{H}\right)
		\end{align}
		\label{eq: hermite}
	\end{subequations}
	for each mode $n$.
	Here, $h = c_s^2 \ln(\rho/\rho_0)$ is the enthalpy, taking the place of pressure and density in the isothermal disc.
	For $n=1$, the horizontal velocity normal modes are linearly increasing with height, whereas the vertical velocity amplitude is constant.
	This vertical structure of the perturbations and background stratification means that the perturbation energy quickly falls off away from the mid-plane.
	It is hypothesised that because the stratification constrains the perturbation energy to be localised near the mid-plane, the non-linear development of the parametric instability will be significantly different than that found in \citet{Papaloizou:2005fo}.
	In the absence of strong vertical stratification, such as in the Earth's atmosphere and as approximated in \citet{Papaloizou:2005fl}, the perturbation energy remains nearly constant in~$z$.

	\subsubsection{Parametric instability of inertial waves}
	\label{sec: parametricInstability}
		
		Analogous to the eccentric parametric instability producing eccentricity growth in the global disc equations,
		parametric resonance may be excited in the local model by the inherent modulation of the physical parameters including the orbital convergence of streamlines as well as the tidal potential, $\Phi_2$, at a frequency $\Omega$, due to the prescribed eccentricity.
		However, now the parametric driving frequency is set as $\Omega$, and the natural oscillation frequencies of the inertial waves are selected, given the resonance condition
		\begin{equation}
			\omega =  \frac{m}{2} \Omega.
		\end{equation}
		The strongest ($m=1$) resonance occurs then for inertial waves with frequency $\omega = \sfrac{1}{2} \Omega_0$,
		which corresponds to the wavenumber $k_\xi H = \sfrac{1}{2}\sqrt{3(4n-1)} = 3/2$.
			
		The founding linear instability analysis studying the growth of this parametric instability in constant eccentricity discs was carried out by \citet{Papaloizou:2005fl}.
		His local stability analysis applies only near the mid-plane where he makes the approximation of a cylindrical disc gravitational potential, $\Phi = -GM/r$, independent of $z$, such that there is no vertical stratification.
		In the limit that $n, k_\xi \to \infty$ it was found that the oscillating divergence of the non-circular streamlines around the disc produces a growth rate of $\sigma = \frac{3}{16} e$.
		In follow-up non-linear global simulations, \citet{Papaloizou:2005fo} additionally found small-scale subsonic turbulence as the instability saturates, leading to a slow decay of eccentricity.
		\citetalias{BarkerOgilvie:2014} extended this initial linear theory to account for the vertical structure of the disc due to a spherical gravitational potential varying off of the mid-plane.
		They found that the effect of the enhanced vertical oscillations made possible by the vertical stratification contributes an additional free energy source to the parametric growth.
		Consequently a larger growth rate of $\sigma = \frac{3}{4} e$ was shown, compared to that with no vertical structure.
		This linear theory inevitably breaks down as the inertial waves grow too large. At that point, non-linear effects become important, eventually causing the inertial waves to break.

	\subsubsection{Inertial wave breaking}
	\label{sec: waveBreaking}
		
		A criterion for the critical amplitude at which these parametrically excited inertial waves break will now be constructed, first by order of magnitude kinematic arguments, and then more physically grounded in the stability of rotating fluids.
		
		Generically, waves are seen breaking as their constituent particle velocities exceed the phase velocity, leading to steepening gradients.
		A similar argument will now be made for inertial waves.
		Restricting the wavevector to be in the $\mathbf{\hat{\xi}}$ direction, the inertial wave perturbations are then described by
		\begin{equation}
			\tilde{v} = \hat{v} \mathrm{e}^{\mathrm{i}(k_\xi \xi - \omega t)}.
		\end{equation}
		The phase speed of this perturbation is $v_p = \omega / k_\xi$,
		which suggests that the critical velocity for breaking is $\hat{v}^\xi_\mathrm{crit} \approx \omega / k_\xi$.
		
		A more physically-motivated argument may be constructed by suggesting that inertial waves break when they locally violate the Rayleigh stability criterion --
		where the angular momentum gradient locally becomes negative,
		\begin{equation}
			\frac{\mathrm{d}\ell}{\mathrm{d}r} = \frac{\mathrm{d}}{\mathrm{d}r}\left( r^2 |v^\phi| \right) < 0.
		\end{equation}
		The angular momentum equation for collisionless particles in the local model is
		\begin{equation}
			\ell = R^2 \tilde{v}^\eta + 2\Omega R \xi  
		\end{equation}
		where $\tilde{v}^\eta$ is the angular velocity perturbation from the local eccentric model.
		Equating the radial derivative to $0$, and using the fact that for inertial perturbations
		\begin{equation}
			\tilde{v}^\xi = \frac{1}{2}\frac{R\omega}{\Omega} \tilde{v}^\eta,
		\end{equation}
		then we obtain the same result that
		\begin{equation}
			\hat{v}^\xi_\mathrm{crit} = \omega / k_\xi.
			\label{eq: breakingThreshold}
		\end{equation}
		Written in terms of the particle displacement, $\delta \xi$,
		this shows that breaking occurs when $k_\xi \delta \xi \approx 1$, i.e. neighbouring particles crash into each other.
		
		Finally, the critical height above the mid-plane that the inertial waves will begin to break, is found to be
		\begin{equation}
			z_\mathrm{crit} = \frac{\omega}{k_\xi \hat{v}_0^\xi} \mathrm{e}^{-\sigma t} 
		\end{equation}
		for the $n=1$ mode of the profile given by (\ref{eq: hermite}) with initial amplitude $\hat{v}_0^\xi$ at $z = 1$. 
		Thus looking far enough off of the mid-plane, the inertial waves can always be seen to break.
		Without considering non-linear effects, then as the parametric instability grows, the breaking region approaches the mid-plane if unbalanced by some means of transporting energy in $k$-space.
		
		Although these criteria are motivated in a circular disc, they are nonetheless found to be a good approximation for small eccentricities when compared to numerical simulations.

\section{Numerical method and setup}
\label{sec: numericalMethod}

An overall second accurate order finite volume scheme is employed to evolve the governing system of equations in conservation form. To maintain this accuracy, however, it is necessary to exactly maintain the laminar dynamic equilibrium solutions by construction.
Thus, a number of further approximations and modifications must be made to the local quasi-axisymmetric model (\ref{eq: localEccentric}) to make the numerical implementation tractable.
Nonetheless, the numerical model developed here will be presented for general thermodynamics and disc geometry ($e’$ \& $\omega’$);
however, the results will focus solely on the effect of varying the eccentricity.


Unfortunately, the eccentricity looks like an $m=1$ azimuthal inertial mode, and so a component of $\partial_\xi p$ remains in equation (\ref{eq: etaVelocity}) when attempting to reduce to axisymmetry.
However, by writing equation (\ref{eq: etaVelocity}) rather for the \textit{covariant} velocity,
\begin{equation}
	v_\eta = g_{\lambda\phi}v^\xi + g_{\phi\phi}v^\eta,
\end{equation}
the (now angular momentum-like) equation becomes properly axisymmetric.
This is akin to writing the azimuthal momentum equation in terms of the specific angular momentum, which is a conserved quantity for a general eccentric disc, unlike the angular velocity.		
This becomes a problem only in eccentric discs due to the variation of the radial distance from the centre along an orbit.
Thus the evolution of the covariant $v_\eta$ replacing (\ref{eq: etaVelocity}) is
\begin{equation}
	Dv_\eta 
			= \left(g_{\phi\phi}\Omega_\lambda + \Omega \left( g_{\phi\phi}\Gamma^\phi_{\lambda\phi} - g_{\lambda\lambda}\Gamma^\lambda_{\phi\phi} - g_{\lambda\phi}\Gamma^\phi_{\phi\phi}  \right) \right)v^\xi  +  g_{\phi\phi}\Omega_\phi v^\eta,
	\label{eq: newEtaVelocity}
\end{equation}
reducing (\ref{eq: localEccentric}) to a $2.5$-dimensional model. 


A further modification to the local eccentric model of \citetalias{OgilvieBarker:2014} will also be made to help formulate a more well-posed vertical boundary condition, as well as permit the construction of a well-balanced numerical method.
By transforming the vertical $z$ coordinate into one following the vertical ``breathing'' of the laminar solution, then these oscillations simply become a static equilibrium.
In this oscillating frame, reflective boundary conditions exactly coincide with the laminar solution, and so it becomes unnecessary to introduce ``non-physical'' fluid inflowing near the boundaries just to maintain a coherent vertical structure.
Additionally, with a proper spatial reconstruction and integration scheme, the laminar equilibrium solution can be maintained to machine precision.

The governing equations (\ref{eq: localEccentric}) along with (\ref{eq: newEtaVelocity}) are thus transformed into the oscillatory $\zeta$ coordinate, described by $z \to \zeta H(t)$.
This transformation modifies the following relevant geometric terms from the definitions in the orbital coordinates of \citet{Ogilvie:2001ja}:
\begin{subequations}
	\begin{align}
		&\partial_z \to H^{-1} \partial_\zeta \\
		&\partial_t|_z \to \partial_t|_\zeta - \frac{\dot{H}}{H}\zeta \partial_\zeta \\
		&v^z \to H v^\zeta + \dot{H}\zeta = H v^\zeta + \bar{v}^z \\
		&g^{\zeta\zeta} = \frac{1}{H^2} \\
		&J' = RR_\lambda H \\
		&\Delta' = \Delta + \frac{\dot{H}}{H}.
	\end{align}
	\label{eq: transformation}
\end{subequations}
In particular, the orbital divergence term, $\Delta'$, picks up a contribution from the vertical laminar oscillations.

The conservation form of the governing equations rewritten in the $(\xi,\eta, \zeta)$ coordinates are then
\begin{equation}
\partial_t
		\begin{pmatrix}
		
		\rho \\
		\rho v^\xi \\
		\rho v_\eta \\
		\rho v^\zeta \\
		\tilde{E}
		
		\end{pmatrix}
	+ \partial_\xi
		\begin{pmatrix}
		
		\rho v^\xi \\
		\rho v^\xi v^\xi + g^{\lambda\lambda}p \\
		\rho v^\xi v_\eta \\
		\rho v^\xi v^\zeta \\
		\left(\tilde{E}+p \right) v^\xi
		
		\end{pmatrix}
	+ \partial_\zeta
		\begin{pmatrix}
		
		\rho v^\zeta \\
		\rho v^\xi v^\zeta \\
		\rho v_\eta v^\zeta \\
		\rho v^\zeta v^\zeta + g^{\zeta\zeta}p \\
		\left(\tilde{E}+p \right) v^\zeta
		
		\end{pmatrix}
= 	\mathbf{S}(\mathbf{U}),
	\label{eq: localEccentricVariableZConservative}
\end{equation}
where the perturbation energy density, $\tilde{E}$, must be defined as 
\begin{equation}
	\tilde{E} = \rho \epsilon + \sfrac{1}{2}\rho \left( g_{\lambda\lambda}v^\xi v^\xi + 2g_{\lambda\phi}v^\xi v^\eta + g_{\phi\phi}v^\eta v^\eta + g_{\zeta\zeta}v^\zeta v^\zeta \right),
\end{equation}
to cancel out the additional vertical pressure gradient term in the energy equation.
The source terms on the right side of each of the equations are respectively
\begin{subequations}
	\begin{align}
		S_\rho =& - \rho\Delta' \\
		S_\xi =& - \rho v^\xi \Delta' - 2\rho\Omega \left(\Gamma^\lambda_{\lambda\phi} v^\xi + \Gamma^\lambda_{\phi\phi} v^\eta\right)  \\
		S_\eta =& - \rho v_\eta \left(\Delta' + \Omega_\phi\right) + \\ 
				& \; \rho v^\xi \left( g_{\lambda\phi}\Omega_\phi - g_{\phi\phi}\Omega_\lambda + \left(g_{\lambda\lambda}\Gamma^\lambda_{\phi\phi}+g_{\lambda\phi}\Gamma^\phi_{\phi\phi}-g_{\phi\phi}\Gamma^\phi_{\lambda\phi}\right)\Omega \right) \nonumber \\
		S_\zeta =& - \rho v^\zeta \Delta' - \rho\Phi_2 \zeta  - 2 \rho v^\zeta \frac{\dot{H}}{H} - \rho \frac{\ddot{H}}{H} \zeta  \\
		S_E =& - \left(\tilde{E}+p\right)\Delta' - \rho \Omega v_\eta \left(v^\xi \Gamma^\phi_{\lambda\phi} + v^\eta \Gamma^\phi_{\phi\phi}\right) + \\
				& \; \rho\Phi_2 H^2 \left( \frac{1}{2}\zeta^2\Delta - \zeta v^\zeta \right) - \rho v^\zeta H \left( \dot{H} v^\zeta + \ddot{H}\zeta \right). \nonumber
	\end{align}
	\label{eq: sourceVariable}
\end{subequations}
%
%
%
%
%
%
Notably, two additional source terms arise in the $\zeta$-momentum equation as a result of this transformation into vertically oscillating coordinates.
The first comes from the flux of $\zeta$-momentum generated by the grid velocity, and the second term which includes $\ddot{H}$ arises due to the acceleration of the grid in this even further non-inertial coordinate system.

\subsection{Modified finite volume scheme}
\label{sec: modifiedFVM}
	
	With these simplifying approximations, the 2.5-dimensional system (\ref{eq: localEccentricVariableZConservative}) is solved with an overall second order finite volume scheme.
	The scheme achieves this second order accuracy via a slope-limited MUSCL-type spatial reconstruction which accommodates the now-static pressure base state \citep{Kappeli:2016gg}.
	It is further dimensionally split, and employs an exact Riemann solver.
	%
	However, the $g^{\lambda\lambda}$ and $g^{\zeta\zeta}$ inverse metric terms that arise in the $\xi$ and $\zeta$ flux now imply that the finite volume advection is anisotropic.
	As these metric terms vary with time, the $\xi$ and $\zeta$ coordinates stretch and contract periodically around the disc. Thus the characteristics in an inertial cartesian system where the Riemann problem is posed are curved, which requires consideration of the generalised Riemann problem.
	Nonetheless, to a good approximation, the geometric coefficients may be held constant for the short duration of each advection time step. Convergence can then be shown as $\Delta t \to 0$, as the piecewise-linear characteristics better represent the true curved characteristics.
	In this approximation, the standard Riemann problem is transformed by simply scaling the characteristics by $\sqrt{g^{\lambda\lambda}}$ and $\sqrt{g^{\zeta\zeta}}$ when solving in the split $\xi$ and $\zeta$ direction, respectively.
	However, this approximation would be expected to break down near pericentre for large eccentricity discs.
	
	
	From the local expansion in \S \ref{sec: localEccentric}, each of the geometric terms are evaluated on the reference orbit, $(\lambda_0, \varphi(t),0)$, and so are implicit functions of time only.
	The orbital location of the local coordinates, $\varphi(t)$, must then be determined as a function of time.
	This requires numerically solving the ODE,
	\begin{equation}
		\frac{\mathrm{d}\varphi}{\mathrm{d}t} = \Omega(\lambda_0,\varphi) = \sqrt{\frac{GM}{\lambda_0^3}}\left(1+e\cos\theta\right)^2,
	\end{equation}
	here by Kepler's method with Newton-Raphson iterations to obtain a very accurate solution.
	The transformation into $\zeta$ coordinates (\ref{eq: transformation}) also requires the horizontally invariant laminar solution, $H(t)$, described by equation (\ref{eq: odeH}) to be found at each time step.
	For efficiency, each of the metric components, including $H(t)$, are computed only once at the beginning of each simulation, and then these tabulated values used around each consecutive orbit.

\subsection{Well-balanced correction}
\label{sec: wellBalancedCorrection}
	
\begin{figure}
		\begin{center}
    		\includegraphics[scale=1]{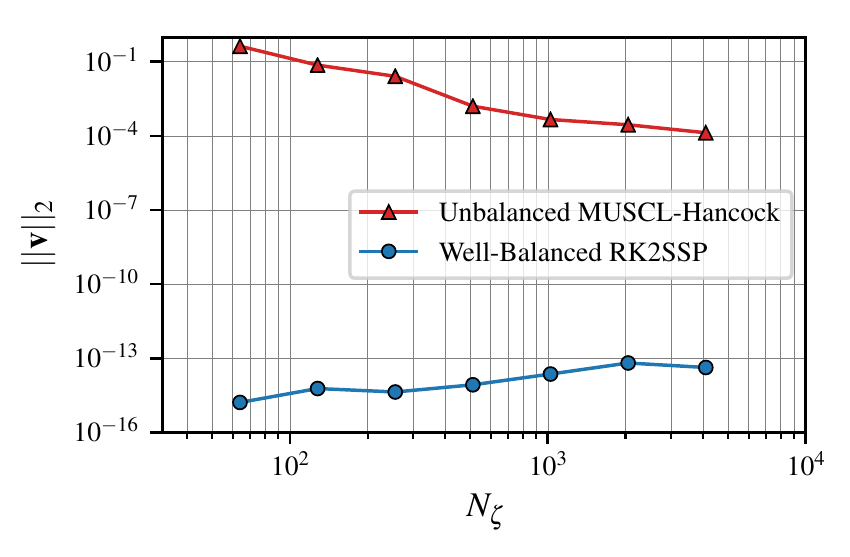}
    		\caption{The $L_2$-norm error of the velocity for the well-balanced and unbalanced schemes, 
						in a test to hold the ``static'' laminar solution over $10$ orbits of a disc with $e=0.1$.}
			\label{fig: equilibriumError}
  		\end{center}
	\end{figure}

	A cursory implementation of the traditional source splitting method produces significant numerical artefacts on the seemingly trivial problem of maintaining the vertical hydrostatic equilibrium solution \citep[as pointed out in][]{Slyz:1999hs}.
	This is because the hydrostatic pressure profile, which at cell interfaces manifests as a pressure jump, is interpreted by the Riemann solver as a propagating acoustic wave in the opposite direction of the potential gradient. Even with second order parallel source splitting, the advection terms are not exactly balanced with the source term, as it is analytically in 
	\begin{equation}
		\frac{\partial p}{\partial z} = - \rho \frac{\partial \Phi}{\partial z}.
		\label{eq: hydrostaticEquilibrium}
	\end{equation}
	This results in an unacceptably large accumulation of errors as apparent in Fig.~\ref{fig: equilibriumError}.
	
	The transformation into $\zeta$ coordinates (\ref{eq: transformation}) permits the implementation of so-called ``well-balanced'' schemes \citep{Greenberg:1996bv} which exactly satisfy a discrete form of this equilibrium.
	Most implementations assume some underlying thermodynamic equilibrium to accomplish this \citep{LeVeque:1997eg,Botta:2004cd,Chandrashekar:2015hq}. But for our purposes, we implement a method similar to that introduced by \citet{Kappeli:2016gg} which relies on a MUSCL-type reconstruction of both the global pressure profile and the perturbation quantities using only the local potential gradient.
	For this purpose, the additional $-\rho \ddot{H}/H \zeta$ source term in (\ref{eq: sourceVariable}) should be interpreted as coming from a dynamic potential, $\ddot{H}/H$, and must be included into the global potential, $\Phi$, in order for the well-balanced reconstruction to maintain this ``static'' equilibrium.
	
	This well-balanced method finally requires that the gradient source terms on the RHS of (\ref{eq: localEccentricVariableZConservative}) be solved unsplit from the vertical advection integration scheme including the pressure terms physically balancing the equilibrium.
	To this end, a second order Runge-Kutta strong stability preserving (RK2SSP) scheme \citep{Gottlieb:2001iy} integrates the advection terms unsplit with the equilibrium source terms,
	\begin{equation}
		S_\mathrm{e} = -\rho \left( \Phi_2 + \frac{\ddot{H}}{H} \right).
	\end{equation}
	The remaining source components are then integrated by parallel operator splitting using an adaptive step size Runge-Kutta (Fehlberg) $4^\mathrm{th}$ order with a $5^\mathrm{th}$ order error estimator.

\subsection{Problem setup}
\label{sec: problemSetup}
	
	
	The numerical implementation of the methods described above, as well as the following results, are given in 
	units of length such that the circular disc isothermal scale height, $H_0$, is unity.
	The unit of time is also defined such that the angular velocity $\Omega_0$ at $r=\lambda_0$ is unity, so that the linear velocities are non-dimensionalised by the isothermal sound speed. 
	Finally, the mid-plane density at apocentre is set to be $1$.
	
	
	The standard numerical problem setup used for each of the base runs evolves an initial perturbation on a vertically-centred domain with height $L_\zeta = 12$, and width $L_\xi$ chosen to exactly fit the wavelength of the perturbation, $2\pi/k_\xi$.
	Although tests included the $n=1,2, \, \& \, 3$ inertial modes, the following will focus on the strongest ($n=1$) mode because it is expected to dominate the saturation characteristics.
	For $n = 1$, the resonant $\omega = 1/2$ perturbation has a radial wavenumber $k_\xi = 3/2$.
	For this mode, the vast majority of the perturbation energy is represented in the domain, with less than $10^{-7}$ of the total energy not captured in the first $6$ scale heights.
	The domain is initialised with the discrete isothermal density profile for a dynamic neutrally stratified disc, and the equilibrium velocity set to naught.
	An exact inertial wave perturbation is added to produce a standing wave with $|v^\xi| = 10^{-3}$.
	The form of this perturbation is
	\begin{equation}
		\begin{pmatrix}
			
			h' \\
			v^\xi \\
			v^\eta \\
			v^\zeta
			
		\end{pmatrix}
		=
		\begin{pmatrix}
			
			-k_\xi \omega^2 \\
			\omega (\omega^2-n) \\
			\frac{1}{2}(\omega^2-n) \\
			n k_\xi \omega/H
			
		\end{pmatrix}
		\cdot
		\begin{pmatrix}
			
			\sin(\omega t + \sfrac{\pi}{4}) \; \sin(k_\xi \xi - \sfrac{\pi}{4}) \\
			\sin(\omega t - \sfrac{\pi}{4}) \; \sin(k_\xi \xi + \sfrac{\pi}{4}) \\
			\sin(\omega t + \sfrac{\pi}{4}) \; \sin(k_\xi \xi + \sfrac{\pi}{4}) \\
			\sin(\omega t - \sfrac{\pi}{4}) \; \sin(k_\xi \xi - \sfrac{\pi}{4})
			
		\end{pmatrix}
		\label{eq: exactPert}
	\end{equation}
	which has a vertical structure given by (\ref{eq: hermite}).
	
	
	A standard resolution of $N_\xi = 128$ is used, with $N_\zeta$ chosen so that $\Delta \xi = \Delta \zeta$ to minimise any anisotropic numerical diffusion that might produce spurious results. For $n=1$ then, $N_\zeta = 366$.
	The Minbee limiter is chosen for these standard runs because, although more numerically diffusive, it is necessary for stability near the vertical boundaries for larger eccentricities.
	With this resolution and a CFL number of $0.9$, a typical evolution time of $1000$ orbital periods requires around $500$ CPU-hours.

\subsection{Validation of code}
\label{sec: validationOfCode}
	
	We now present an extensive series of validation tests conducted to build confidence in the results of this new code.
	We start with the finite volume advection scheme and source integration for a cartesian metric, ensuring this implementation passes the simple one- and two-dimensional test suite designed by \citet{Toro:1999by}. 
	As expected, compared with the often-used MUSCL-Hancock scheme, the RK2SSP integrator necessitated by the well-balanced correction is found to be slightly more dissipative, smearing out strong shocks particularly in Toro's modified Sod shock tube (test $1$).
	A number of tests specific to this implementation as it pertains to the local disc model will now be presented.
	
	\subsubsection{One-dimensional laminar solutions}
	
		We now present tests in the untransformed $z$ coordinates and ensure agreement with the semi-analytic horizontally invariant laminar solution described by (\ref{eq: odeH}).
		The results over a single orbit for eccentricities up to $0.5$ show good agreement with the analytic solution;
		however, for longer evolution times with moderate eccentricities, the vertical boundaries begin to compound errors by slowly degrading the similarity solution profile (\ref{eq: primitivesForm}), and eventually breaking down completely.
		Even worse, the no-flux boundaries produce a weak rarefaction wave which eventually impacts the mid-plane within a few sound-crossing times. Further attempts with a transmissive reconstructive boundary still result in an ill-posed problem due to non-physical fluid inflow.
		Hence, the oscillatory vertical geometry ($\zeta$) described by (\ref{eq: transformation}) will now be adopted and tested.
				
		The second order RK2SSP well-balanced implementation is compared against the standard unbalanced MUSCL-Hancock scheme in a convergence study of the deviation from the laminar (now static) equilibrium.
		This discrete equilibrium with no initial velocity perturbation is evolved for $10$ periods in a disc with $e=0.1$ so as to also demonstrate the consistency of the quasi-constant geometry approximation.
		As evidenced in Fig.~\ref{fig: equilibriumError}, the well-balanced scheme nearly maintains the laminar oscillations to machine precision.
		Although near the mid-plane it maintains the equilibrium exactly to machine precision,
		near the boundaries the machine precision perturbations from the mid-plane amplify as they travel into the increasingly rarefied atmosphere.
		This is contrasted with the unbalanced MUSCL-Hancock scheme, in which errors arise due to the operationally-split nature of the algorithm trying to balance the pressure gradient independently from the potential gradient in the source term.
		Thus without the well-balanced method, a prohibitive $10^8$ grid points would be required with this second order source-split scheme.

	\subsubsection{Circular disc}
	
		
		It is also crucial that this new implementation, in particular with the modified source and dimensional splitting method, produces the correct physical frequencies in response to specific perturbations.
		Thus the initial conditions (\ref{eq: exactPert}) with $e=0$ and $n = 0,1, \, \& \, 2$ are evolved for $1000$ orbital periods to gain high frequency resolution.
		The natural oscillation frequencies are extracted and shown as black x's in Fig.~\ref{fig: dispersion}. Both the acoustic and inertial branches are captured to within the tight frequency resolution of the test.
		
		
		Throughout the course of these long runs the amplitude of the initial perturbation inevitably decays.
		This effective $\alpha$-viscosity due to numerical dissipation may be measured given the properties of the test perturbation and the decay rate by
		\begin{equation}
			\alpha \sim \frac{\sigma_\mathrm{damp}}{k_\xi^2 + n}.
		\end{equation}
		For a typical run with $k_\xi = 3/2$ and $n=1$, the perturbation decays to nearly $50$ per cent of the initial amplitude over the course of $1000$ periods, corresponding to a damping rate of $10^{-4}$.
		The effective numerical $\alpha$-viscosity is then approximately $\alpha_\mathrm{n} \approx 3\cdot 10^{-5}$.
		Although this exceeds the typical molecular viscosity $\alpha \approx 10^{-9}$, it is still orders of magnitude smaller than the expected and observed values of $\alpha$, as well as more than $100$ times less diffusive than comparable global simulations with vertical structure \citep{Bitsch:2013}.

	\subsubsection{Eccentric disc linear growth}
		
		\begin{table}
			\centering
			\begin{tabular}{cccccc}
				\hline
				$n$  &  $e$  &  $\sigma_\mathrm{linear}$  &  $\sigma_\mathrm{Floquet}$  &  $\sigma_\mathrm{numerical}$   &  $\alpha_\mathrm{eff}$ \\
				\hline
				1  &   0.01  &  0.0075  &  0.007498  &   0.007301   &   6.065e-5   \\
				1  &   0.03  &  0.0225  &  0.02245   &   0.02221    &   7.297e-5   \\
				1  &   0.05  &  0.0375  &  0.03728   &   0.03689    &   1.199e-4   \\
				1  &   0.1   &  0.075   &  0.07335   &   0.07147    &   5.788e-4   \\
				1  &   0.3   &  0.225   &  0.1993    &   0.1712     &   8.640e-3   \\
				\hline
				2  &   0.01  &  0.0075  &  0.007499  &   0.007183   &   4.367e-5   \\
				2  &   0.03  &  0.0225  &  0.02249   &   0.02214    &   4.797e-5   \\
				2  &   0.05  &  0.0375  &  0.03743   &   0.03698    &   6.274e-5   \\
				2  &   0.1   &  0.075   &  0.07440   &   0.07301    &   1.927e-4   \\
				\hline
			\end{tabular}
			\caption{Energy growth rate validation in the linear regime. The growth rates calculated from the linear theory for small $e$, from Floquet theory \citepalias{BarkerOgilvie:2014}, and from the current non-linear numerical simulations are compared, along with the effective $\alpha$-viscosity attributed to the numerical dissipation.}
			\label{tab: LinearGrowth}
		\end{table}
		
		The linear growth rate of these simulations will now be shown to agree with the $\sigma = (3/4)e$ from linear theory, less this numerical damping rate.
		We set a non-zero eccentricity, and initialise each simulation with the exact inertial mode perturbation at a specific energy of $10^{-10}$, well within the linear regime.
		The resulting growth rates of these tests for the standard problem setup of \S \ref{sec: problemSetup} are given in Table~\ref{tab: LinearGrowth}.
		Although the dissipation corresponds with an effective $\alpha$ of at least $10^{-4}$, the deviation from the linear relation in $e$ for large eccentricity follows the Floquet results of \citepalias{BarkerOgilvie:2014}.
		Further, because the errors in the linear regime are attributed to numerical dissipation of the methods used, they are shown to decrease with increasing resolution.
		The $n=2$ resonant mode is also verified to ensure that even non-linear wave-wave interactions which may excite the $n=2$ modes can still be captured with good precision.

	\subsubsection{Convergence study}
		
		\begin{table*}
			\centering
			\begin{tabular}{cccccccc}
				\hline
				 $e$  &  $N_\xi\times N_\zeta$  &  $L_\xi/\lambda$  &  $\tau_\mathrm{end}$  &  $\varepsilon_{k,\mathrm{hi}}$  &  $\varepsilon_{k,\mathrm{lo}}$  & $\alpha_\mathrm{sat}$  &  $\partial_\tau e_\mathrm{sat}$\\
				\hline
				0.01 & 128x366 & 1 & 700 & 0.01354 & 0.008194 & -4.316e-6 & -4.769e-6 \\
				0.01 & 256x732 & 1 & 300 & 0.01679 & 0.006500 & -1.038e-5 & -6.134e-6 \\
				\hline
				0.02 & 128x366 & 1 & 700 & 0.03886 & 0.01076 & -5.430e-5 & -1.599e-5 \\
				0.02 & 256x732 & 1 & 300 & 0.03606 & 0.005928 & -3.892e-5 & -1.319e-5 \\
				0.02 & 384x1098 & 3 & 450 & 0.03543 & 0.008434 & -3.953e-5 & -1.433e-5 \\
				\hline
				0.03 & 128x366 & 1 & 700 & 0.05769 & 0.01885 & -1.850e-4 & -2.310e-5 \\
				0.03 & 256x732 & 1 & 300 & 0.05297 & 0.005666 & -1.520e-4 & -2.013e-5 \\
				0.03 & 640x1830 & 5 & 350 & 0.05210 & 0.009778 & -1.349e-4 & -2.057e-5 \\
				\hline
				0.04 & 128x366 & 1 & 600 & 0.08134 & --- & -3.284e-4 & -3.068e-5 \\
				0.04 & 256x732 & 1 & 250 & 0.09475 & --- & -4.803e-4 & -3.454e-5 \\
				0.04 & 384x1098 & 3 & 400 & 0.07989 & --- & -3.069e-4 & -2.985e-5 \\
				0.04 & 640x1830 & 5 & 200 & 0.07923 & --- & -3.022e-4 & -2.946e-5 \\
				\hline
				0.05 & 128x366 & 1 & 600 & 0.09104 & --- & -4.395e-4 & -3.151e-5 \\
				0.05 & 256x732 & 1 & 250 & 0.1002 & --- & -5.935e-4 & -3.467e-5 \\
				0.05 & 384x1098 & 3 & 400 & 0.09344 & --- & -3.738e-4 & -2.944e-5 \\
				0.05 & 640x1830 & 5 & 200 & 0.09222 & --- & -3.646e-4 & -2.905e-5 \\
				\hline
				0.1 & 128x366 & 1 & 500 & 0.1158 & --- & -1.130e-3 & -3.879e-5 \\
				0.1 & 256x732 & 1 & 200 & 0.1243 & --- & -1.076e-3 & -3.646e-5 \\
				0.1 & 640x1830 & 5 & 200 & 0.1289 & --- & -1.017e-3 & -3.816e-5 \\
				\hline
				0.2 & 128x366 & 1 & 350 & 0.2234 & --- & -1.012e-3 & -5.590e-5 \\
				0.2 & 256x732 & 1 & 200 & 0.2333 & --- & -1.377e-3 & -5.862e-5 \\
				\hline
			\end{tabular}
			\caption{Saturation characteristics from the resolution and domain size ($L_\xi$) convergence study. Conditional averages based on the top $20$ per cent of the specific kinetic energy are used to compute $\varepsilon_{k,\mathrm{hi}}$, $\alpha_\mathrm{sat}$, and $\partial_\tau e_\mathrm{sat}$ at a more representative saturation value. Similarly, a conditional average on the bottom $80$ per cent represents $\varepsilon_{k,\mathrm{lo}}$ in the lower energy state.}
			\label{tab: nonlinearTable}
		\end{table*}
		
		A comprehensive convergence study was performed to ensure that the following presented numerical experiments correctly resolve the necessary physics and are independent of the choices made in \S \ref{sec: problemSetup} for the resolution, initial and boundary conditions, time-stepping, and domain size.
		The results of this convergence study are summarised in Table \ref{tab: nonlinearTable}.
		
		
		Convergence is first demonstrated with increasing resolution (or equivalently, with decreasing numerical dissipation).
		We find that the energy saturation level is insensitive to further increasing resolution;
		however, second-order statistics such as the Reynolds stresses are notoriously difficult to converge without sub-grid turbulence models or direct numerical simulations (DNS).
		This is because although often insignificant to the larger scales (given no significant backscatter interaction), subgrid-scale fluctuations still contribute to the Reynolds stresses, and therefore $\alpha$ and the eccentricity decay rate. This means that the value calculated for $\alpha$ is inevitably dependent on the grid size \citep{King:2007kh}.
		Indeed \citet{Fromang:2007jj} and \citet{Johansen:2009} have reported a roughly linear dependence of the transport coefficients on resolution, likely continuing until $\Delta x$ approaches the Kolmogorov scale.
		Thus as expected, with $4$ times higher resolution, the magnitudes computed for the Reynolds stresses consistently increase by around $20$ per cent.
		The computed $\alpha$ is then lower than the true value in magnitude solely because the energy spectrum is truncated by the finite grid size.
		This effect consequently impacts the convergence of the observed non-linear behaviour at saturation.
		At saturation, weakly non-linear oscillations arise from the cyclic generation and dissipation of large-scale zonal flows.
		It is not surprising then that the period of the limit cycles produced for intermediate eccentricities has not yet converged.
		This is because the slow dynamics dictating the limit cycle time-scale are governed by the turbulent mixing (diffusion) time,
		$\tau_\mathrm{mix} \sim 1/(k_\xi^2 \nu_\mathrm{tot})$, where the total viscosity, $\nu_\mathrm{tot}$, has contributions from the turbulent as well as numerical viscosities.
		As the turbulence previously generated during resonance decays, $\nu_T$ decreases, and so eventually the numerical viscosity dominates the dissipation.
		Nonetheless, the limit cycle size and shape in phase space are still converged.
		It should still be pointed out that there is varied statistical convergence in the high-resolution convergence runs due to shorter evolution times.
		In certain runs with long limit cycles, only the upper saturation level of the first cycle are captured, and so these statistics are not fully converged, but is believed to be due to sampling.
		
		
		We now investigate the impact of the limited domain size on the saturation characteristics.
		By repeating the base simulation in an increased radial domain size of $2$, $3$, and $5$ times $\lambda_\xi$ with equivalent resolution, it is confirmed that the most unstable mode and its saturation are faithfully represented by a $1 \lambda_\xi$ radial box.
		The radial profile of the zonal structures are additionally unaffected by the constrained domain, and continue to prefer the $2 k_\xi$ wavenumber.
		It could have been expected that because higher order unstable modes are less biased against in large domains, that these modes may become active and contribute to the kinetic energy at saturation. While a spatio-temporal frequency analysis indeed shows an increased presence of the $n=2$ mode at saturation, the resulting turbulent kinetic energy is still largely independent of the radial domain size.
		
		
		
		
		Chaotic behaviour is common for non-linear dynamical systems, such as this self-regulating instability.
		This behaviour is well-known to cause extreme sensitivity to the initial conditions,
		yet must still be statistically independent of the initial conditions for long times (and at saturation).
		To confirm this, we initialise the base setup with a random white-noise velocity perturbation.
		The resonant $n=1$ mode inevitably biased by the choice of the domain size, $L_\xi = 4\pi/3$, eventually establishes the correct linear growth rate, and consistently saturates at the same energy level.

\section{Non-linear saturation}
\label{sec: nonlinearSaturation}

\subsection{Development of breaking inertial waves}
\label{sec: breakingWaves}
	
	The well-balanced finite volume method evolving the local eccentric 2.5-dimensional model as described previously has been implemented and will now be employed to study the non-linear effects which are initially produced by the breaking inertial waves.
	The breakdown of inertial waves is the crucial step in the energy pathway from the parametrically unstable inertial mode feeding into small-scale turbulence, and further radial transport.
	
\begin{figure}
		\includegraphics[scale=1]{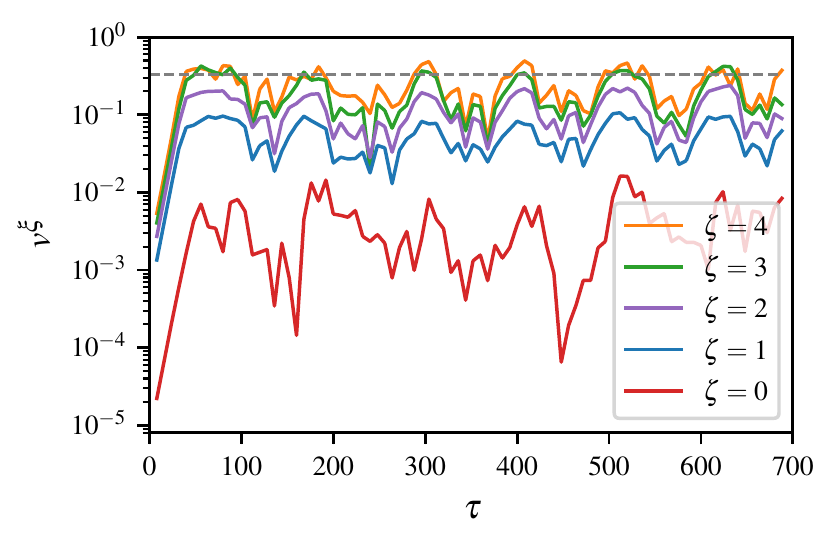}
		\caption{The velocity of the parametrically unstable ($\omega=1/2$) mode is shown at various heights in the $e=0.03$ disc, where each height is averaged over $2$ oscillation periods (i.e. $4$ orbits). The inertial waves are unstable for $\zeta \gtrsim 3$, where $v^\xi$ oscillates about the critical inertial wave breaking amplitude, $v^\xi_\mathrm{crit} = 1/3$ (dotted line). }
		\label{fig: amplitudeRegulate}
	\end{figure}

	The initial perturbation introduced is well within the linear growth regime of this parametric instability.
	However, this linear theory soon begins to break down and the linear growth is saturated.
	This is initially brought on by the breaking of inertial waves, but when fully developed includes more complex wave-wave interactions which could additionally excite higher-order azimuthal modes in a fully three-dimensional model.
	In an unrestricted vertical disc, due to the monotonically increasing ($n=1$) vertical profile (\ref{eq: hermite}), there always exists a height above the mid-plane with a large enough amplitude to feel these non-linear effects, however small their contribution to the total energy.
	In a limited computational domain, however, only eventually will the waves begin to break at the vertical extents, which first appears as a localised spike in the azimuthal shear.
	The criterion for breaking from \S \ref{sec: waveBreaking} that $k_\xi \delta \xi \approx 1$ implies that at criticality, there is a local increased convergence from the fluid over-extension in $\mathbf{\hat{\xi}}$, explaining the consequent vertical jet impinging towards the mid-plane.
	Fig.~\ref{fig: amplitudeRegulate} further seems to agree with the theoretical critical amplitude, $v^\xi_\mathrm{crit} = 1/3$. The parametrically unstable component of the radial velocity appears to oscillate very near to this critical value while at the upper saturation value (in the absence of zonal flows), suggesting that the additional energy contributes to the turbulence.
	
\subsection{Heuristic energy transport model}
\label{sec: heuristic}
	
\begin{figure}
		\includegraphics[scale=1]{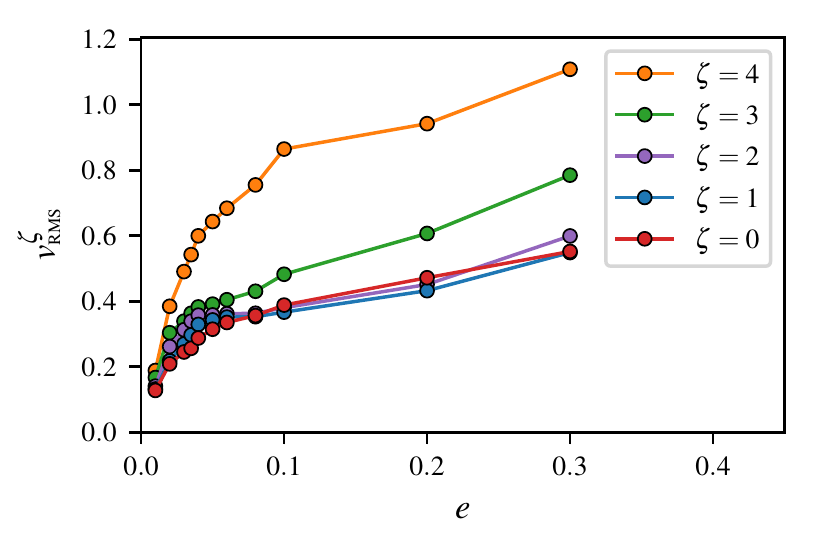}
		\caption{The vertical root-mean-square perturbation velocity is shown at various heights scaling with the eccentricity. The majority of the motion remains subsonic, though with large enough forcing the vertical extents of the disc become supersonic.}
		\label{fig: RMSvelocity}
	\end{figure}

	
	After this initial ordered wave breaking period and a slight overshoot, the disc relaxes into a quasi-steady turbulent state.
	This turbulence is characterised by subsonic fluctuations with a typical root-mean-square vertical velocity of $v^\zeta_{\textsc{RMS}} \approx 0.4 c_s$, except at large distances from the mid-plane for moderate eccentricity where the fluctuations are marginally supersonic.
	This dependence on eccentricity is shown in Fig.~\ref{fig: RMSvelocity} and may be compared to the typical velocities an order of magnitude smaller that were found in the numerical simulations by \citet{Papaloizou:2005fo} in a disc without vertical structure.
	His global disc simulations with $e \approx 0.13$ produced typical $v^z_{\textsc{RMS}} \approx 0.03 c_s$ constant over the height of the disc.
	
	The influence this vertical structure has on the nature of this turbulence is quite evident from Fig.~\ref{fig: amplitudeRegulate}.
	Only the parametrically excited inertial waves above $\zeta > 3$ are ever allowed to grow large enough in amplitude to begin breaking down and cascading energy to smaller scales.
	At a steady state, then this localised energy transfer out of the parametrically unstable mode suggests energy transport within the $\omega = 1/2$ mode from $\zeta \lesssim 3$ to higher $\zeta$.
	
\begin{figure}
		\includegraphics[scale=1]{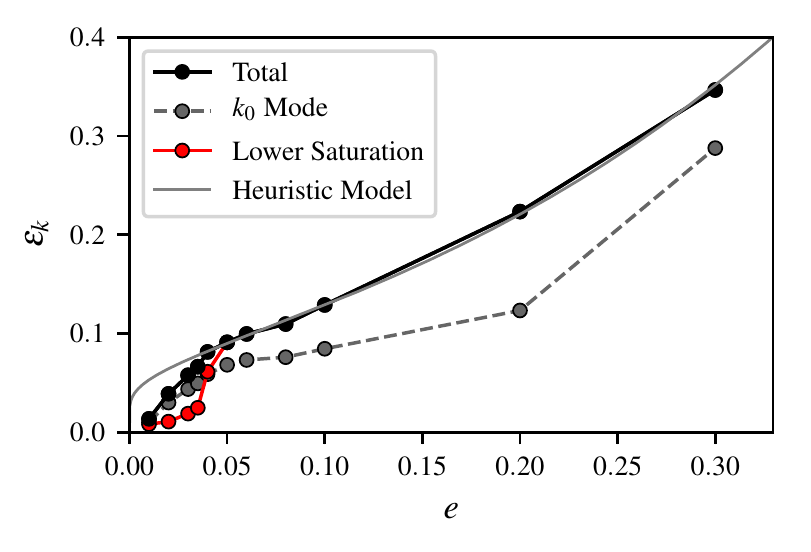}
		\caption{The total specific turbulent kinetic energy at saturation is shown compared to the contribution of the energy contained in the $k_0$ parametrically excited mode. The lower saturation energy level (conditionally averaged on the bottom $80$ per cent) in the zonal dissipation regime of the limit cycle is also shown when it significantly differs from the mean at small eccentricities as evident in Fig.~\ref{fig: Eoscillate}. The energy heuristic (\ref{eq: EsatScaling}) is also fitted to this parameter sweep in $e$.}
		\label{fig: saturateKEfit}
	\end{figure}

	
	A heuristic model may then be constructed by using turbulent kinetic energy transfer and localisation arguments to gain insight into the energy saturation scaling observed in Fig.~\ref{fig: saturateKEfit}.
	We will now detail the energy pathway as it is transferred from
		the disc eccentricity 
		into the parametrically unstable inertial mode, 
		then vertically into the upper atmosphere of the disc 
		where it is able to grow to large enough amplitude, 
		break down into turbulence, 
		and cascade down to smaller scales 
		where viscous dissipation finally takes its toll.
	For the purposes of finding the natural energy saturation level, we make an energy balance vertically over the entire column,
	where we must now assume statistically time-homogeneous turbulence in spectral equilibrium.
	
	Energy is injected by the disc eccentricity at the parametrically resonant wavenumber, $k_0$, at a rate proportional to the \textit{local} specific kinetic energy of the unstable mode defined as
	\begin{equation}
		\varepsilon_{k,0}(\zeta) \equiv \varepsilon_k (k_0,\zeta) = \frac{1}{2}\frac{\left< \rho g_{ij}v^i v^j \right>(k_0,\zeta)}{\left< \rho \right>}.
	\end{equation}
	Here $\left< \cdot \right>$ represents averages over a horizontal plane at a given height off the mid-plane, such that the total kinetic energy contained in the half-disc may be written as
	\begin{equation}
		\mathcal{E} = \int_0^\infty \left< \rho \right> \varepsilon_k \; \mathrm{d}\zeta.
		\label{eq: defE}
	\end{equation}
	Disregarding the effect of zonal flows so that the energy locally sourced by the resonance is directly proportional to $\varepsilon_{k,0}$, then the total rate of kinetic energy sourced into the disc is
	\begin{equation}
		\partial_t \mathcal{E}_\mathrm{in} = \int_0^\infty \frac{3}{2}e\Omega \left<\rho \right> \varepsilon_{k,0} \, \mathrm{d}\zeta \sim \mathcal{E} e \Omega.
	\end{equation}
	The approximation that the majority of the kinetic energy is stored in the unstable inertial mode ($\mathcal{E}_0 \sim \mathcal{E}$) has been used here, and is shown in Fig.~\ref{fig: saturateKEfit} to have similar scaling.
	The important effect that zonal flows have on the saturation dynamics will be considered only in \S \ref{sec: largeScaleZonalFlows}, but for now will simply be averaged over by assuming a steady-state.
	This modal energy consequently propagates upwards into the disc atmosphere, through $\zeta_\mathrm{crit}$ where non-linear effects break down and further generate a turbulent cascade down to the Kolmogorov length-scale (or numerically, the grid scale).
	The total rate of energy removal from the domain is thus approximated as
	\begin{equation}
		\partial_t \mathcal{E}_\mathrm{out} \approx \int_{\zeta_\mathrm{crit}}^\infty \left<\rho \right> \nu_T k_0^2 \varepsilon_k \, \mathrm{d}\zeta,
		\label{eq: Eloss}
	\end{equation}
	where $\nu_T$ is the turbulent viscosity.
	A \textit{local} scaling for this turbulent viscosity motivated by the mixing length model for large eddy transport suggests
		$\nu_T \propto \sqrt{\varepsilon_k} / k_0$.
	Additionally, the linear perturbation velocity profile ($n=1$) is assumed for all $\zeta$, so that
	\begin{equation}
		\varepsilon_k = \frac{\varepsilon_{k,\mathrm{crit}}}{\zeta_\mathrm{crit}^2} \zeta^2,
	\end{equation}
	allowing the scaling of integral (\ref{eq: Eloss}) to be found.
	A similar analysis may also be done for the other inertial modes using profiles of higher powers in $\zeta$, with conceptually similar results.
	
	Balancing these two energy integrals finally gives a scaling for the total energy at saturation,
	\begin{equation}
		\mathcal{E} = \frac{a}{e} \frac{2+\zeta_\mathrm{crit}^2}{\zeta_\mathrm{crit}^3} \mathrm{e}^{-\zeta_\mathrm{crit}^2/2} 
	\end{equation}
	for some proportionality constant $a$.
	To find closure, the total disc energy (\ref{eq: defE}) must be approximated using $\zeta_\mathrm{crit}$ by
	\begin{equation}
		\mathcal{E} \approx \frac{\varepsilon_{k,\mathrm{crit}}}{\zeta_\mathrm{crit}^2} \int_0^\infty \left< \rho \right> \zeta^2 \, \mathrm{d}\zeta \approx \frac{b}{\zeta_\mathrm{crit}^2}.
	\end{equation}
	where it was assumed that $\varepsilon_{k,\mathrm{crit}} \propto \left(v^\xi_\mathrm{crit} \right)^2 \propto 1$ as suggested by \S \ref{sec: waveBreaking}.
		The magnitude of the proportionality constant $b$ is then expected to be $b \approx 3\sqrt{\pi/2} \rho_0 \left(v^\xi_\mathrm{crit} \right)^2$.
	The total disc energy can then be implicitly represented in terms of the eccentricity as
	\begin{equation}
		\frac{a^2\left( b + 2 \mathcal{E} \right)^2 }{b^3 \mathcal{E}} \mathrm{e}^{-b/\mathcal{E}} = e^2,
		\label{eq: EsatScaling}
	\end{equation}
	which may be numerically solved for $\mathcal{E}$.
	
	
	The best-fitting function of the form (\ref{eq: EsatScaling}) is found for the constants $a=0.114$ and $b=0.574$, and is plotted over the saturation energy scaling in Fig.~\ref{fig: saturateKEfit}.
	This value of $b$ implies that $\varepsilon_{k,\mathrm{crit}} \approx 0.46$, or equivalently that $v^\xi_\mathrm{crit} \approx 0.39$.
	This critical value for $v^\xi$ is only slightly larger than the value of $0.33$ given by equation (\ref{eq: breakingThreshold}), but could be expected as the argument presented there was only indicating the point of initial breaking.
	
	\begin{figure}
		\includegraphics[scale=1]{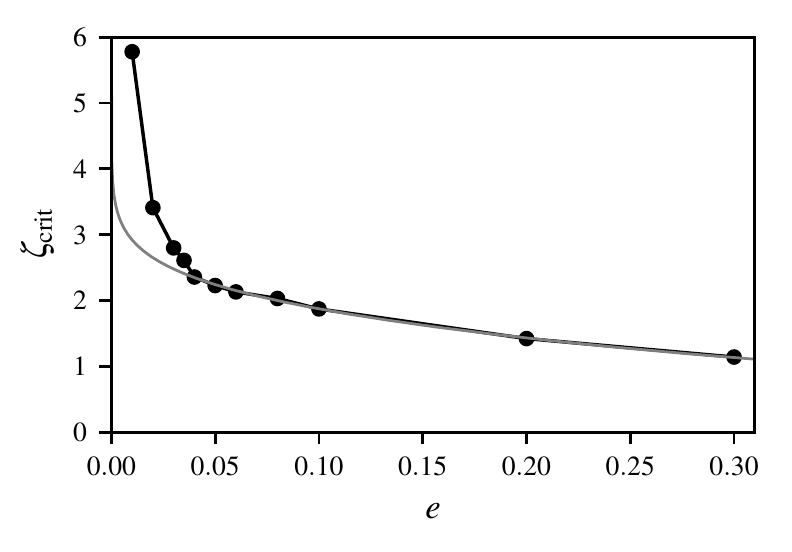}
		\caption{The predicted saturation height above the disc mid-plane for the heuristic model compared to in simulations, assuming $v^\xi_\mathrm{crit}$ is constant. This represents the height at which breaking inertial waves dissipate the most energy.}
		\label{fig: saturateZeta}
	\end{figure}
	
	The values for $\zeta_\mathrm{crit}$ predicted by this model are compared with those found numerically and are shown in Fig.~\ref{fig: saturateZeta}.
	As more energy is required to be dissipated (with increasing eccentricity), then $\zeta_\mathrm{crit}$ must encroach on the higher densities near the mid-plane.
	Thus, only localised parts of the disc both actively support inertial wave breaking \emph{and} are able to dissipate sufficient energy.
	Below $\zeta_\mathrm{crit}$, the inertial waves are not significantly non-linear, whereas well above $\zeta_\mathrm{crit}$ the perturbation energy available to dissipate rapidly decreases in the more rarefied atmosphere.
	It may then be expected that $\zeta_\mathrm{crit}$ remains further from the mid-plane for higher inertial modes because the perturbation energy profile extends much higher.
	Nonetheless, as it was shown that $v^\xi_\mathrm{crit}$ is nearly constant, the most kinetic energy is available to the $n=1$ mode because this profile is contained nearest the mid-plane.
	This particular behaviour is a consequence of the vertical disc structure, and so is unique to stratified atmospheres.
	
	
	The effects of higher-order azimuthal modes are also ignored in this energetics argument, as they are not supported in our 2.5-dimensional model.
	Although the locally axisymmetric mode indeed dominates the linear regime, it is expected that high-order azimuthal modes would additionally be excited in the strongly non-linear regime.
	These modes would likely store and help dissipate additional energy, but at the same time introduce only a small source of energy from the weaker parametric resonance \citepalias{BarkerOgilvie:2014}.
	We therefore expect that the preceding energetics arguments and thus the non-linear results for energy saturation (as well as eccentricity decay rate) to be only slightly modified.
	
	
	Disagreement of this heuristic model at small eccentricities may be explained by the presence of large-scale zonal flows which reduce, or even completely turn off, the parametric energy source.
	The saturated turbulence is thus not statistically time-stationary,
	and so a conditional average (of the top $20$ per cent of the turbulent kinetic energy) is required to extract only the upper saturation level which corresponds to times with minimal zonal flow.
	This effect of zonal flows is weakened for large eccentricity, and so the approximation to disregard them holds up better.
	Here, the forced turbulence becomes sufficiently strong to disrupt these coherent flows, and the broad frequency-bandwidth for resonance (see fig.~1 in \citepalias{BarkerOgilvie:2014}) reduces their ability to completely mitigate the primary instability.

\subsection{Angular momentum transport}
\label{sec: alphaViscosity}
	
\begin{figure}
		\includegraphics[scale=1]{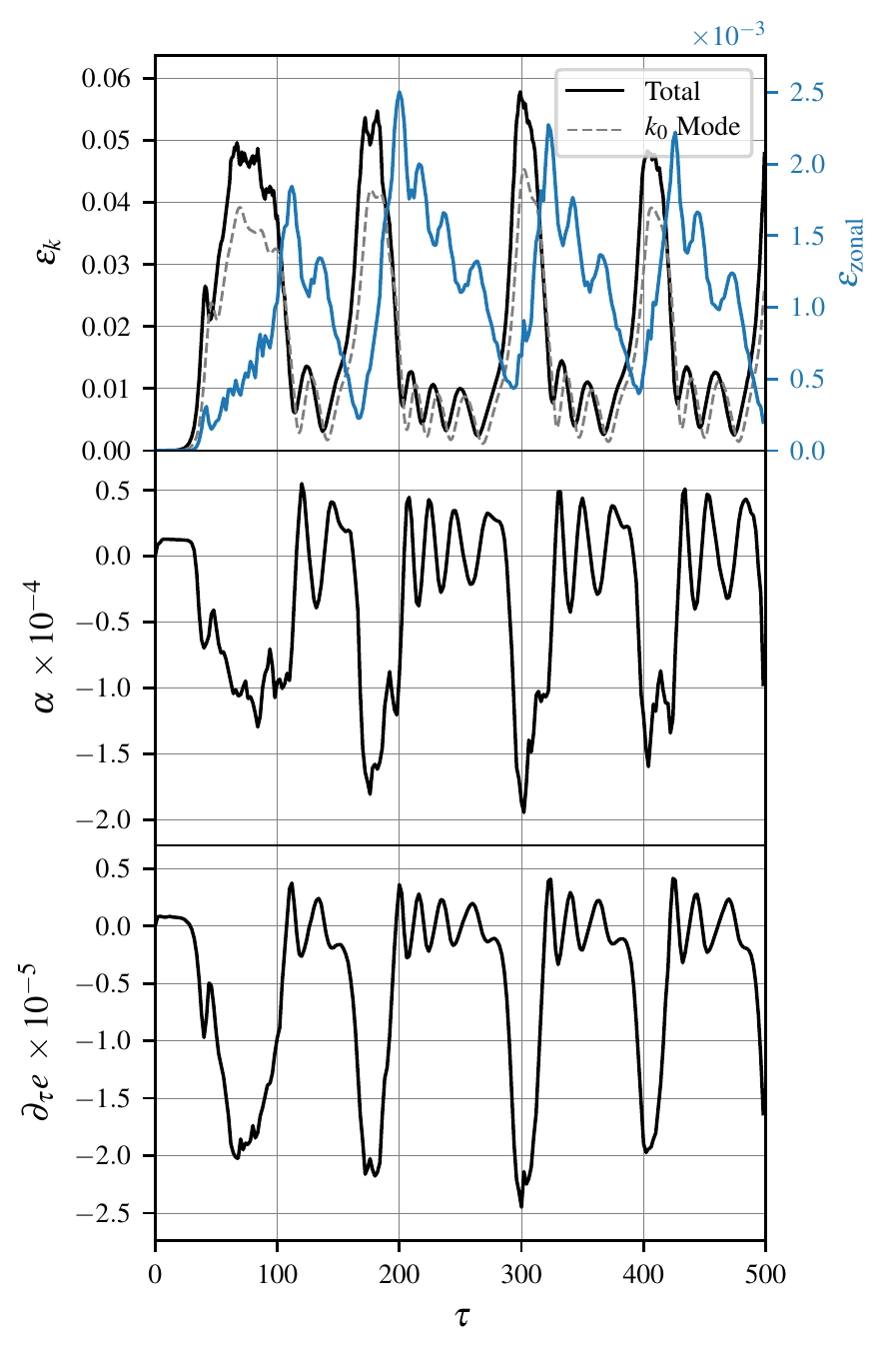}
		\caption{Typical weakly non-linear oscillations connecting the evolution of specific perturbation and zonal energies (top), the $\alpha$-viscosity (middle), and the eccentricity decay rate (bottom) for the base $e=0.03$ disc. The energy contained in the parametrically unstable $k_0$ mode is also shown on top (dashed). The dissipation-type quantities, $\alpha$ and $\partial_\tau e$, are closely related, and anti-correlated to the energy saturation level.}
		\label{fig: compositePlot}
	\end{figure}

	
	The integral (\ref{eq: alphaViscosity}) over the disc is compiled discretely within the code after each time step to determine the efficiency with which this turbulent state transports angular momentum.
	Most instabilities in discs extract energy from the background differential rotation (shear).
	This instability rather extracts energy from the background eccentric oscillations, and so does not immediately guarantee outward angular momentum transport ($\alpha > 0$) on average.
	A typical time evolution of $\alpha$ is shown in Fig.~\ref{fig: compositePlot}, illustrating this contention with two distinct phases.
	The conditionally averaged $\alpha$ at the high turbulent energy saturation indicates a negative $\alpha$ on average. 
	On the contrary, a slightly positive $\alpha$ on average exists during passive periods of zonal flow decay and relatively small turbulence intensity.
	
	
	Axisymmetric instabilities are known to produce very little angular momentum transport, such that the turbulent viscosity mixing down the angular momentum gradient dominates in $\alpha$.
	Each fluid parcel conserves its initial angular momentum, $\ell \propto \sqrt{\lambda}$, and so as the turbulence rearranges the fluid it also effectively transports angular momentum inwards, producing $\alpha \lesssim 0$.
	In a similar two-dimensional axisymmetric model of tidally distorted discs, \citet{Ryu:1994dh} also found negligible angular momentum transport (albeit there being quite dependent on the initial conditions).
	Small negative $\alpha$ are also produced, for example, in weak three-dimensional convective instabilities, where the shear restricts the evolution to be effectively two-dimensional axisymmetric \citep{Lesur2010}.
	While our 2.5-dimensional model preserves the three-dimensional nature of the saturated turbulence (particularly by allowing the vortex-stretching term, $(\mathbf{\omega}\cdot \mathbf{\nabla})\cdot\mathbf{u}$), the axisymmetry approximation may still be artificially constraining the angular momentum transport.
	Therefore, present results are expected to be most valid in three-dimensional discs when the Keplerian shear effectively constrains the non-linear sourcing of broad-band turbulence to be two-dimensional, which is expected for discs of moderate eccentricity.
	It is speculated then for a three-dimensional local model with sufficiently large forcing (eccentricity), that the turbulence could overcome the tendency towards 2-dimensionality by the shear, and allow $\alpha > 0$.
	This would be of interest to investigate in future three-dimensional eccentric shearing box simulations, 
	and compare with the work of \citet{Papaloizou:2005fo} who found an $\alpha \sim 10^{-3}$ for $e=0.1$.
	
	
	A rough scaling for $\alpha$ may still be found by rewriting the energy dissipation rate as 
	\begin{equation}
		\partial_t \mathcal{E}_\mathrm{out} \approx \frac{\alpha}{\Omega} \int_{\zeta_\mathrm{crit}}^\infty \left<p\right> k_0^2 \varepsilon_k \, \mathrm{d}\zeta,
	\end{equation}
	and then equating with  (\ref{eq: Eloss}).
	As expected, this gives $\alpha \propto \left<\sqrt{\varepsilon_k}\right>$.
	In the averaged sense, then $\alpha \propto \mathcal{E}$, suggesting that $\alpha$ should indeed continue to grow with eccentricity.
	
\subsection{Eccentricity evolution}
\label{sec: eccentricityEvolution}
	
	\begin{figure}
		\includegraphics[scale=1]{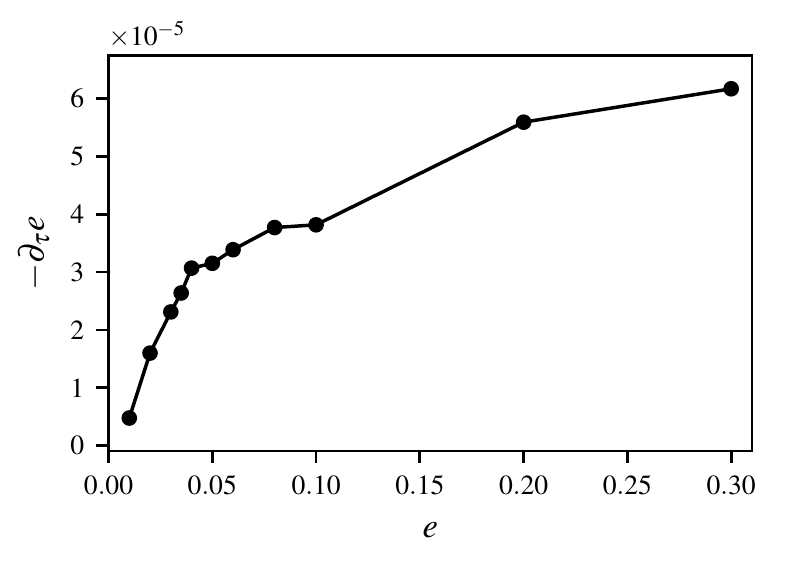}
		\caption{The conditionally averaged eccentricity decay rate at saturation is shown scaling with the disc eccentricity.
		For comparison, the eccentricity growth rate by the $3$:$1$ eccentric Lindblad resonance found by \citet{Lubow:2010fv} is more than $100$ times faster than this damping.} 
		\label{fig: saturateDE}
	\end{figure}
	
	The integral (\ref{eq: eccentricityEvolution}) over the disc is also computed at run-time, and assumes the radial structure of a stationary alpha-disc with constant $\alpha(r)$. The aspect ratio is taken to be $H_0/\lambda_0 = 0.02$, which is typical for cataclysmic variables, and selected for comparison to \citet{Lubow:2010fv}.
	A typical time evolution of $\partial_\tau e$ is also displayed in Fig.~\ref{fig: compositePlot}, which shows that it is similarly inflicted by the non-linear oscillations and also precedes the zonal flows by $\pi/2$.
	Nonetheless, it is found that if the parametric instability is active and saturated, any disc eccentricity decays within about a thousand orbital periods.
	
	Assuming a constant alpha-disc model, \citet{Lubow:2010fv} found an eccentricity growth rate driven by the eccentric Lindblad resonance of
	\begin{equation}
		\partial_\tau e \simeq \frac{2\pi e}{10\mathcal{P}_b},
	\end{equation}
	where $\mathcal{P}_b$ is the binary orbit period relative to the local disc period.
	%
	The eccentricity growth rate by the $3$:$1$ eccentric inner Lindblad resonance (thought to be active in superhump binaries) is more than $100$ times larger than the eccentricity decay rate found in this numerical model (shown in Fig.~\ref{fig: saturateDE}). 
	In reality, the non-linear evolution of this instability is also likely to modify the assumed global disc structure with increasing eccentricity. 
	So although these two phenomena would be unable to balance and produce a stable eccentricity, the results do suggest ``bursty'' \emph{global} dynamics reminiscent of superoutbursts and superhumps.
	%
	While the parametric instability is inhibited by zonal flows, the disc remains quiescent until sufficient energy has been stored in the disc eccentricity.
	As the zonal flows decay, the reactivated parametric instability once again opens access to this stored energy allowing the eccentricity to decay and dissipating energy within a thousand orbits. 
	This limit-cycle behaviour will now be explored further in the following section.

\section{Large-scale zonal flows}
\label{sec: largeScaleZonalFlows}

\subsection{Background}
\label{sec: zonalBackground}

\begin{figure}
		\includegraphics[scale=1]{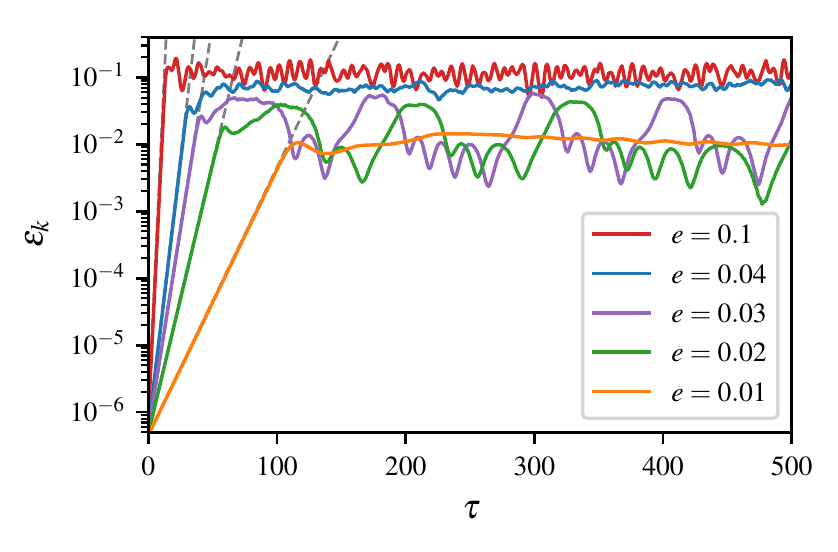}
		\caption{Comparison of the long-time dynamic saturation behaviour with increasing eccentricity, showing the first few limit cycles until $\tau = 500$ for representative base runs. Low frequency weakly non-linear oscillations are evident in particular for $e=0.02$ \& $0.03$. Theoretical linear growth is indicated by grey dashed lines.}
		\label{fig: Eoscillate}
	\end{figure}

	
	Weakly non-linear oscillations are not uncommon in saturated systems sourced by a property-dependent linear growth.
	As the basic state is modified, inhibiting this linear growth, the previous equilibrium system begins to decay.
	Large-scale, azimuthally-symmetric ($m=0$) azimuthal momentum fluctuations with finite $k_\xi$ are exactly this saturation mechanism modifying the basic state, preventing further parametric energy sourcing into inertial waves, and pre-empting the wave-breaking saturation mechanism.
	This globally axisymmetric zonal mode sees the local flow alternating with radius through sub- and super-Keplerian features.
	Without these zonal flows, a statistically time-stationary state would be achieved when the parametric instability energy sourcing directly balances the decay of the resultant turbulent kinetic energy, as assumed in the simplified heuristic in \S \ref{sec: nonlinearSaturation}. However, much richer behaviour is observed (as seen in Fig.~\ref{fig: Eoscillate}), the details of which depend strongly on the eccentricity.
	Still, the presence of these oscillations are found to be quite robust for intermediate eccentricities; however, their exact quality is inevitably sensitive to the numerical dissipation.
	
	Much literature exists in the plasma physics community where zonal flows are known to be crucial in regulating drift-wave turbulence and suppressing instabilities.
	Large-scale structures generated out of drift-wave turbulence arising during magnetic confinement in tokamaks has long been known, and was first explained by \citet{Diamond:1991ci}.
	A good review of these developments regarding zonal flows in plasmas is given by \citet{Diamond:2005br}.
	
	Zonal flows are also prevalent in geophysical contexts, such as in planetary boundary layers, where they are attributable to the jet streams and tidal currents (albeit meandering) on Earth.
	The original scaling arguments describing these zonal flows in two-dimensional turbulence were formulated by \citet{Rhines:1975}.
	Further work by \citet{Sun:1993kl} show these flows arising out of convective turbulence, and more recently by \citet{Barker:2016jla} for the elliptic instability in tidally-deformed planets.
	A review of zonal flow observations and potential generation mechanisms in the Jovian atmosphere \citep{Vasavada:2005gs}, as well as a thorough analogy between magnetised plasmas and quasi-geostrophic fluids \citep{Gurcan:2015jy} are good resources.
	
	In the context of astrophysical discs, it has been known for some time that the forced two-dimensional turbulence generated by the magneto-rotational instability (MRI) is capable of generating strong zonal fields and flows via the Lorentz force \citep{Johansen:2009,Kunz:2013jp}.
	More recently, however, a purely hydrodynamic viscous overstability mechanism has been identified by \citet{Vanon:2017bd} to produce growing azimuthal velocities and potentially develop into zonal flows.

\subsection{Zonal flow dynamics}
\label{sec: zonalFlowDynamics}
	
	\subsubsection{Driving zonal flows}
	\label{sec: drivingZonalFlows}
		
		
		We now hope to develop a model to describe some of this oscillatory behaviour found at saturation.
		Qualitatively, these zonal flows are driven by non-linear mode coupling producing large-scale variations in the transport coefficients.
		Thus non-local interactions between the zonal flow perturbation and finite frequency modes found in the turbulence spectrum allow direct energy transfer to the large-scale stationary mode.
		This process is only similar to the well-known inverse cascade of energy to large scales, but distinctively there via \textit{local} coupling in wavenumber space.
		Our 2.5-dimensional model indeed supports three-dimensional turbulence as shown in \S \ref{sec: alphaViscosity}, and so large-scale eddies still exhibit a forward energy cascade to smaller scales.
		In physical space, this coupling of the turbulence modes to the zonal shear (by way of eddy tilting) is realised through the Reynolds shear stresses; however, the axisymmetry of our numerical model means that there is still no way for pressure gradients to drive the zonal flows.
		
		
		To develop a zonal flow model based on the Reynolds-averaged formulation, we begin by averaging the $v^\eta$-momentum equation (\ref{eq: etaVelocity}) both in $\zeta$ and over the fast time-scale, $t$:
		\begin{equation}
			\left<\cdot\right>(\xi, \tau) \equiv \frac{1}{L_\zeta} \int_{\zeta} \frac{1}{\mathcal{T}} \int_{\tau - \mathcal{T}}^{\tau + \mathcal{T}} \left(\cdot\right) (\xi,\zeta,t') \, \mathrm{d}t \mathrm{d}\zeta,
			\label{eq: averagingOp}
		\end{equation}
		where $\mathcal{T}$ is an intermediate time-scale defining the width of the kernel.
		Exploiting symmetry and homogeneities, the averaged equation may be simplified to
		\begin{equation}
			\partial_t\left< \rho v^\eta \right> + \partial_\xi\left< \rho v^\xi v^\eta \right> = \left< \mathrm{RHS} \right>,
			\label{eq: RANS}
		\end{equation}
		where the averaged RHS term includes all of the linear oscillating and sourcing terms, and is safely dropped a posteriori.
		Because we are interested in a model exhibiting the zonal growth and decay, we take the Boussinesq eddy viscosity approximation to write the Reynolds stress in terms of the large-scale shear,
		\begin{equation}
			T^{\xi\eta} \sim -\nu_T \partial_\xi \left< \rho v^\eta \right>
		\end{equation}
		with the turbulent viscosity approximated as $\nu_T \approx \sqrt{\varepsilon_k}/k$.
		Finally, with large-scale damping, the model equation becomes
		\begin{equation}
			\partial_\tau \left< \rho v^\eta \right> \approx -c_\mathrm{damp} \left< \rho v^\eta \right> + c_\mathrm{eat} \sqrt{\varepsilon_k} \left< \rho v^\eta \right>.
			\label{eq: ppRANS}
		\end{equation}
		Here, $c_\mathrm{eat}$ is the parameterised rate of energy transfer from the turbulent energy into the zonal flows.
		Although these Reynolds stresses in equation (\ref{eq: RANS}) are descriptive as opposed to predictive, various theories predicting their growth have been inspired by a generalised modulational instability with a seed zonal flow, such as the so-called zonostrophic instability \citep {Srinivasan:2012im}, or also from a wave kinetic theory approach \citep{Diamond:2005br}.  
		
		
		\begin{figure}
			\includegraphics[scale=1]{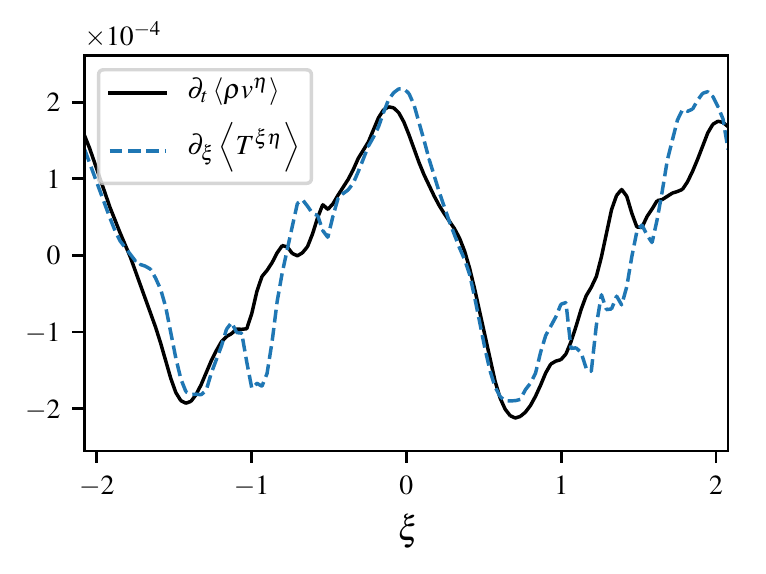}
			\caption{Comparison of the radial structure of the leading order terms in the Reynolds-averaged azimuthal equation (\ref{eq: RANS}) for a disc with $e=0.03$. The robust $2k_0$ structure is evident, in addition to a clear balance indicating that zonal acceleration arises from the asymmetric component of the Reynolds stress gradient.}
			\label{fig: TxyZonalXiForce}
		\end{figure}
		
		We now consider the balance of terms in equation (\ref{eq: RANS}) within the numerical calculations.
		Averaging both the zonal acceleration term and zonal momentum flux term arising from the Reynolds stress over the entire evolution, a strong spatial correlation is evident from Fig.~\ref{fig: TxyZonalXiForce}.
		The zonal flows are observed to consistently take on the preferred $2 k_0$ radial wavenumber compared to the resonant mode.
		It could be suspected that this particular $\xi$ structure might be an artefact biased by the limited and particular domain size, yet the convergence runs summarised in Table~\ref{tab: nonlinearTable} show otherwise.
		Using $2$ to $5$ times the radial box size consistently produces this same zonal wavelength twice that of the parametrically unstable wave.
		In particular cases, an initial zonal flow may still grow with $k_\xi \gtrsim 2 k_0$, yet after a short period an Eckhaus-type instability is observed.
		
		\begin{figure}
			\includegraphics[scale=1]{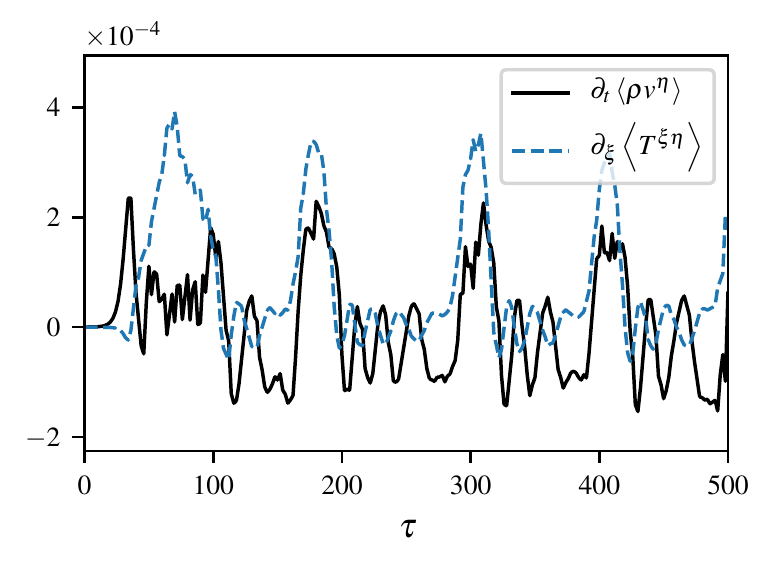}
			\caption{Comparison of the time evolution of the leading order terms in the Reynolds-averaged azimuthal equation (\ref{eq: RANS}). The signals are well-correlated in the fully saturated regime, but less so during the secondary oscillations of zonal flow dissipation.}
			\label{fig: TxyZonalTime}
		\end{figure}
		
		Similarly, a $2$-period time-average is performed to calculate the time-series in Fig.~\ref{fig: TxyZonalTime}.
		The zonal acceleration and momentum flux are better correlated in time in the saturation regime than during the low energy oscillations.
		This suggests that the Reynolds stresses dominate the main limit cycle, yet higher-order terms contained in $\left< \mathrm{RHS} \right>$ may additionally contribute to the secondary oscillations.
		
		\begin{figure}
			\includegraphics[scale=1]{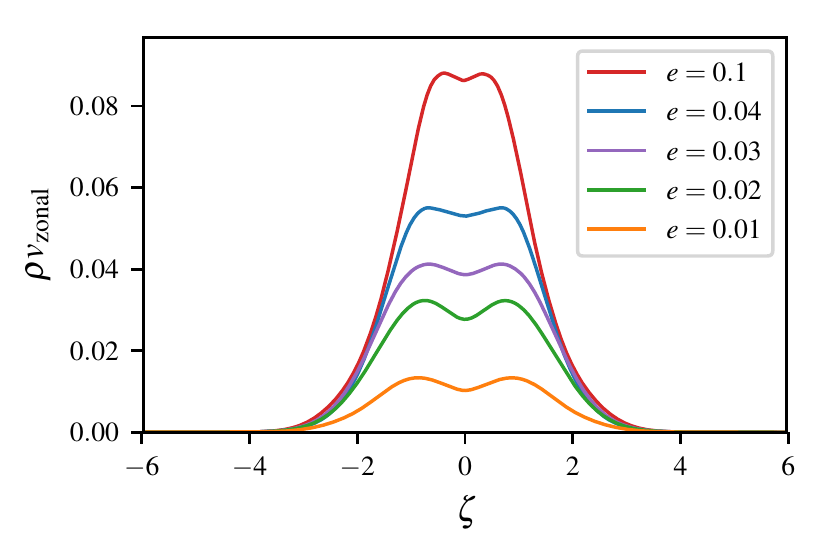}
			\caption{The vertical profile of the contravariant zonal momentum exhibiting a bi-modal structure with each peak centred closer to the mid-plane with increasing forcing amplitude.}
			\label{fig: zonalMomHeight}
		\end{figure}
		
		Finally, the zonal momentum vertical profiles are found to be consistent with the prediction of $\zeta_\mathrm{crit}$ in the heuristic energy model. As displayed in Fig.~\ref{fig: zonalMomHeight}, the zonal momentum is driven to a bi-modal shape, with the peaks decreasing towards the mid-plane with increasing eccentricity. This is because as $\zeta_\mathrm{crit}$ decreases with increasing $e$ (as in Fig.~\ref{fig: saturateZeta}), then the turbulence intensity driving the zonal flows also increases closer to mid-plane.

	\subsubsection{Self-regulation by zonal flows}
	\label{sec: regulatingTurbulence}
		
		
		The growth of these zonal flows are found to be not only mitigated by secondary shear instabilities, but also self-regulated by a feedback loop cutting off their own driving source of energy.
		As these zonal flows grow sufficiently large, they remove access to the free energy source in the eccentricity by locally modifying the 
		inertial frequency of the $k_0$ waves to become incommensurate with the driving frequency.
		The zonal shear contributes to the background vertical vorticity and so modifies the epicyclic frequency to become
		\begin{equation}
			\kappa^2 = 2 \Omega \left( \frac{1}{2} \Omega + R \partial_\xi \left< v^\eta \right> \right) 
			\label{eq: epicyclic}
		\end{equation}
		which consequently shifts the frequency of the $k_0$ inertial wave according to equation (\ref{eq: dispersion}).
		The $k_0$ inertial mode thus ceases to resonate with the eccentric mode, effectively knocking the disc off of resonance. 
		Further, because the new resonant frequency varies on distances less than $2\pi/k_0$ of the inertial wave, no new coherent and growing mode emerges.
		
		\begin{figure}
			\includegraphics[scale=1]{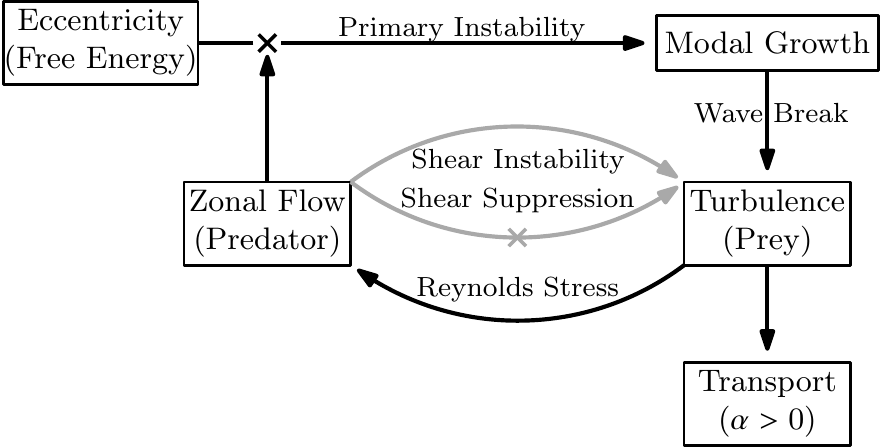}
			\caption{Diagram showing the dominant energy pathway (in black) from the disc eccentricity into turbulent transport, which follows the models proposed in (\ref{eq: ppRANS}) and (\ref{eq: ppTKE}). Secondary effects not modelled are shown in grey.}
			\label{fig: arrowDiagram}
		\end{figure}
		
		With the disc now stable to the parametric instability of inertial waves, the only source of turbulence is removed, and with it the source of driving zonal flows. Any residual turbulent viscosity can now dissipate the zonal flows, allowing the flow to regain resonance, and again grow at near the linear growth rate.
		This process is evident in Fig.~\ref{fig: compositePlot}, though of course because $k_\eta = 0$, the resultant $v^\xi = 0$, and so these zonal flows contribute no net transport.
		We have found that they nonetheless indirectly (and significantly) inhibit transport.
		The resulting dynamics permitting the consistent weakly non-linear oscillations as in Fig.~\ref{fig: Eoscillate} are a direct consequence of this interplay between the primary resonant instability, turbulent dissipation, and zonal flows.
		This flow of energy is summarised in the diagram of Fig.~\ref{fig: arrowDiagram}.
		
		
		To model these oscillations, in addition to the evolution of $\left< \rho v^\eta \right>$ in (\ref{eq: ppRANS}), we must develop a model for the evolution of the saturated turbulent kinetic energy.
		Taking the energy system to be quasi-steady, then the rate of energy extracted from the eccentricity by the primary instability must balance that into $\varepsilon_k$.
		Additionally, the robustness of this instability to being knocked off of resonance is given by the bandwidth of the resonance peak (cf. fig.~1 in \citetalias{BarkerOgilvie:2014} for increasing $e$).
		In that light, we propose a physical model by approximating the resonance peak by a quadratic:
		\begin{equation}
			\partial_\tau \varepsilon_k \approx \sigma_k \varepsilon_k - \frac{\sigma_k}{2 w^2} \varepsilon_k \left< \rho v^\eta \right> ^2.
			\label{eq: ppTKE}
		\end{equation}
		Thus in addition to the energy sourced by the instability slowly turning off with increasing zonal flow, an implicit damping is prescribed as the magnitude exceeds twice the RMS bandwidth, capturing the effect of shear-decorrelation.
		It should also be pointed out from the Floquet analysis of \citetalias{BarkerOgilvie:2014} that not only is the energy growth rate a function of $e$, ($\sigma_k \simeq (3/2) e$), but so is the RMS bandwidth, $w$.
		
		
		Along with the zonal flow model (\ref{eq: ppRANS}), this effective turbulent kinetic energy equation models the processes indicated by the black arrows in Fig.~\ref{fig: arrowDiagram}.
		The zonal shear is also known to directly suppress the existing turbulence by shear-enhanced decorrelation \citep{Biglari:1990hx},
		by refracting the turbulence in $k$-space to smaller scales where it is quickly dissipated \citep{Gurcan:2015jy}.
		This helps to explain the drop off in energy at the peak zonal velocity (see Fig.~\ref{fig: compositePlot}), which far exceeds the turbulent viscous decay rate.
		In direct competition with this mechanism, secondary shear instabilities (such as Kelvin--Helmholtz) preying on the zonal shear profile are found to feed the stored energy back into the turbulence.
		The low-energy oscillations during the zonal decay period (e.g. in Fig.~\ref{fig: Eoscillate}) are attributable to this short-circuiting of the energy pathway.
		Similar sub-oscillations are seen in turbulent Poiseuille flow with zonal structures \citep{Shih:2015dl}, although they have not been explicitly acknowledged.

	\subsubsection{Dependence on forcing magnitude}
	\label{sec: dependenceOnForcing}
		
\begin{figure}
			\includegraphics[scale=1]{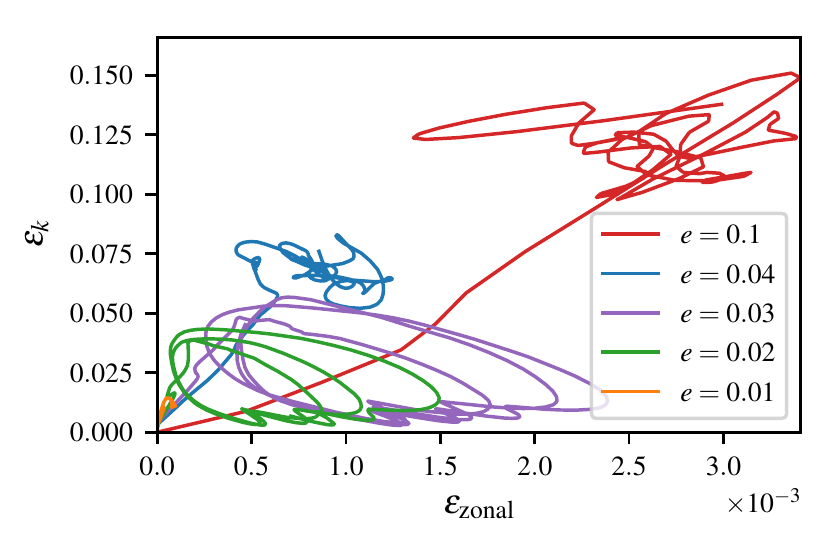}
			\caption{$2$-period temporally-averaged trajectories shown in the zonal and kinetic energy phase space up to $\tau = 400$. Clear limit-cycle behaviour is observed for intermediate eccentricities, whereas statistically-fixed points appear above $e=0.04$.}
			\label{fig: zonalEnergy}
		\end{figure}

		The feedback found above generating the limit-cycle dynamics is found to be quite sensitive to the eccentricity.
		For example, very long time-scale undulations exist at saturation for $e=0.01$, compared to the strongly non-linear saturation at higher energy where eventually there is no obvious separation of time-scales as seen in Fig.~\ref{fig: Eoscillate}.
		The phase portrait showing $\varepsilon_k$ versus $\varepsilon_\mathrm{zonal}$ is given in Fig.~\ref{fig: zonalEnergy}, which makes the effect of zonal flows on the oscillations clear.
		
		We can gain further insight into the changing behaviour with increasing $e$ by a dominant time-scale analysis.
		For small $e \lesssim 0.04$, the frequency bandwidth of instability is very narrow, and so even small zonal flows are able to entirely knock off the resonance.
		Here the limiting time-scale is by the turbulence-enhanced zonal decay,
		\begin{equation}
			\tau_\mathrm{decay} \approx \frac{1}{\nu_T k_0^2} \approx \frac{1}{\sqrt{\varepsilon_k} k_0}.
		\end{equation}
		As usual, $\tau_\mathrm{grow} \approx \sigma^{-1}$, and so as $e$ increases, both time-scales decrease resulting in a higher frequency limit cycle in line with observation.
		
		
		Beyond $e \approx 0.04$, the saturation energy is pushed into a more strongly non-linear regime.
		Although these zonal flows are relatively robust and able to avoid large distortions arising from turbulence,
		past this threshold at $e \approx 0.04$ the zonal flows are no longer able to remain coherent.
		This results in a sudden decay of the zonal flow (rather than growth) when exiting the linear growth regime in Fig.~\ref{fig: zonalEnergy}.
		This sub-critical Hopf bifurcation is attributed to the increasingly broad unstable frequency bandwidth, such that this primary instability is never fully knocked off of resonance.
		Rather, the dynamics attract to a statistically time-stationary state (on a time-scale of a few orbits), which is the result of the balance between the Reynolds stresses driving the zonal flows and the turbulent mixing of the zonal flow.

\subsection{Activator-inhibitor dynamical model}
\label{sec: dynamicalModel}
	
	
	The two effective models including only the most dominant terms in the evolution of the zonal flows (\ref{eq: ppRANS}) and the turbulence kinetic energy (\ref{eq: ppTKE}) form a predator--prey type dynamical system also exhibited in the numerical model (see Fig.~\ref{fig: zonalEnergy}).
	Predator--prey behaviour in turbulent systems was first proposed by \citet{Diamond:1994fd} in the context of plasma drift-wave turbulence.
	We similarly observe this classic behaviour manifest as a competition between coherent structures able to efficiently absorb energy (resonant modes), and those that can withstand the non-linearity for long times (zonal flows).
	Thus interpreting the system as an activator-inhibitor dynamical system will help to gain intuition and develop insight into the otherwise complicated non-linear behaviour.
	
	
	The activator-inhibitor system constructed out of (\ref{eq: ppRANS}) and (\ref{eq: ppTKE}) is
	\begin{subequations}
		\begin{align}
			\partial_\tau v_\mathrm{zonal} &= -c_\mathrm{damp} v_\mathrm{zonal} + c_\mathrm{eat} \sqrt{\varepsilon_k} v_\mathrm{zonal} \\
			\partial_\tau \varepsilon_k &= \sigma_k \varepsilon_k - \frac{\sigma_k}{2 w^2} \varepsilon_k v_\mathrm{zonal} ^2.
		\end{align}
		\label{eq: predatorPrey}
	\end{subequations}
	This may look more familiar after the substitution $\sqrt{\varepsilon_k} = v_\textrm{\textsc{RMS}}$, making clear that the only difference between the classic Lotka-Volterra model is the quadratic zonal flow contribution to the energy predation rate.
	The four stages summarised in Fig.~\ref{fig: arrowDiagram} are:
	\begin{enumerate}
		\item Exponential growth of turbulence (prey) via the primary instability with inertial wave breaking generating broad-band turbulence. The dominant parameter here is $\sigma_k = (3/2) e$.
		\item Activation of the secondary instability zonal flows (predator), feeding off of the turbulence. This predation phase is dictated by the conversion rate, $c_\mathrm{eat}$, of the turbulent velocity fluctuations into the large-scale flows.
		\item Suppression of turbulence and inhibiting the primary instability from continuing to feed energy into the turbulence. This zonal shearing efficiency is parameterised by the bandwidth of instability, $w$.
		\item Damping rate of zonal flows ($c_\mathrm{damp} \approx (3/4) e$) by large scale mixing such as a residual turbulent viscosity. This is the slowest phase, and thus dictates the limit-cycle period.
	\end{enumerate}
	To close the cycle, the linear growth is finally reactivated as soon as the zonal flows are sufficiently damped.
	Of course, although we are unable to capture the sub-critical bifurcation observed near $e = 0.04$, it is proposed that this may be due to a limited ``carrying capacity'' of the turbulence fluctuations in the disc, likely near the sonic point.	
	
	\begin{figure}
		\includegraphics[scale=1]{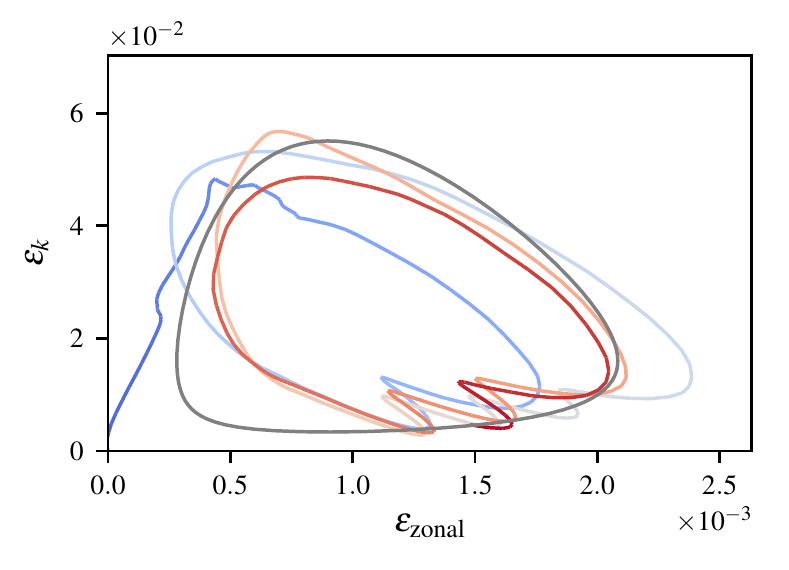}
		\caption{Time evolution of the limit cycle trajectory in the zonal and kinetic energy phase space for $e=0.03$. Changing line colour indicates the time, from dark blue ($\tau = 0$) to dark red ($\tau = 500$). Also shown is the best-fitting curve (grey) based on the predator--prey dynamical model (\ref{eq: predatorPrey}).}
		\label{fig: zonalEnergye03}
	\end{figure}
	
	We compare this model to the limit cycles produced for a disc with $e=0.03$.
	Taking the bandwidth from the Floquet analysis to be $w \approx 0.02$, then the only fitting parameter left is the predation rate.
	Fig.~\ref{fig: zonalEnergye03} shows the time evolution of the $e=0.03$ disc in phase space showing typical limit cycles, along with the best-fitting cycle here with $c_\mathrm{eat} = 0.18$.
	It is apparent that our simplifications indeed prevent the more intricate dynamics such as short-circuit feedback producing these low energy oscillations. 
	Additionally, the deviation from our numerical model as the zonal flows decay seem to indicate more complex zonal decay mechanisms than proposed.
	Nonetheless, the dominant dynamics appear to have been well-captured by our model.


\section{Conclusions}


In this paper, we investigated the saturation of the parametric instability of inertial waves inherent in eccentric discs by way of a new local numerical model, thus extending the linear theory of \citetalias{OgilvieBarker:2014}.
We have constructed the first local numerical model for an eccentric shearing box representing an elliptical accretion disc, thereby generalising the often-used cartesian shearing box model.
We utilised this numerical model to gain understanding of the saturation of this primary instability and resultant non-linear evolution for uniformly eccentric isothermal discs.
The resultant energy pathway from the vertical oscillation modes of the eccentric disc into broad-band turbulence by non-linear inertial wave breaking action is an efficient mechanism for rapid decay of eccentricity.

We have highlighted the importance of the localisation of energy and stresses by the vertical disc structure.
Just as the linear theory for the inertial wave parametric instability produced significantly higher growth rates compared to a cylindrical disc \citep{Papaloizou:2005fl,OgilvieBarker:2014}, we have shown that the non-linear evolution also has interesting differences as the vertical structure localises the effective region of inertial wave breaking to a small region near some critical height off the mid-plane.
Below this height, non-linear effects are found to be very weak; yet far above this region, the density is too low for the dissipation there to contribute significantly.
We have modelled this interplay at saturation by a vertical energy balance (with more success for large eccentricities where the turbulence is statistically time-stationary), shedding light on the observed scaling of the saturation energy with eccentricity in the strongly non-linear regime.

We have also observed the growth of robust, large-scale zonal flows which break the time-homogeneity and dominate the global dynamics by regulating access to the energy stored in the disc eccentricity.
Still, residual turbulent viscosity and secondary shear instabilities damp these zonal structures, setting in motion cyclic behaviour below a critical eccentricity of $e=0.04$.
This ``bursty'' behaviour dictated by the growth and break-down of these zonal structures may explain the occurrence of superoutbursts in SU UMa stars.
We have found that the eccentricity damping rate during the transient saturation periods is around one per cent of the estimated growth by the $3$:$1$ Lindblad resonance in cataclysmic binaries, but may still suggest episodic periods of energy dissipation and eccentricity decay on time-scales of a thousand orbits.
In the context of protoplanetary discs where embedded planets may continually perturb the disc, this eccentricity decay mechanism may have further-reaching effects on the evolution and damping of planet eccentricity \citep{Papaloizou:2001fb,Goldreich:2003,Bitsch:2013}.


The scope of our constructed numerical model has been intentionally simplified, in particular with the implemented thermodynamics, which is deferred to future work.
A non-isothermal equation of state would permit an equilibrium entropy stratification which supports gravity waves, and is expected to impact the energy pathway.
The linear growth rate would be only marginally affected by these gravito-inertial waves because the majority of energy is contained near the mid-plane where the buoyancy frequency is very small.
However, because these waves are limited to propagating between $N < \omega < \kappa$, 
then these waves are confined to $\zeta < \left(2\sqrt{1-1/\gamma}\right)^{-1}$, which for typical values of $\gamma$, is within the first scale height.

There is yet much to be learned about the self-organisation and growth of these large-scale zonal structures, most important being the exact mode coupling which drives the zonal instability.
Future work should also address the question of whether the broken azimuthal symmetry in a fully three-dimensional local model still permits the growth of coherent zonal flows.
In particular, the sensitivity of these inherently axisymmetric zonal flows and the resulting limit cycle feedback mechanism to the presence of high-order azimuthal perturbations which were excluded from our 2.5-dimensional model must be explored.
Although this saturated axisymmetric disc was found to be quite inefficient at transporting angular momentum, more significant outward angular momentum transport could also be expected as the symmetry is broken at large eccentricities. 
The extension of this local eccentric model to three dimensions would require a modified solution of the generalised Riemann problem to allow for off-diagonal pressure terms, as well as a regridding technique in the azimuthal direction to support the expanding and contracting domain. 
Finally, global effects such as disc twist and radial eccentricity gradients were ignored in the presented investigation, although they are known to affect the linear growth of the primary instability, and therefore likely impact the non-linear evolution as well.
It would therefore be quite interesting to couple a model for the disc eccentricity and gradient evolution back into this local model, to observe the effects on the limit cycle behaviour and further support our conclusion regarding outbursts.

\section*{Acknowledgements}

A.F.W. acknowledges support from the National Science Foundation (NSF) Graduate Research Fellowship under grant no. DGE-114747. 
A.F.W. is also grateful for helpful discussions with Adrian Barker and Henrik Latter.
Presented numerical work used an allocation of computer resources on the Stampede supercomputer, awarded by the Extreme Science and Engineering Discovery Environment (XSEDE) program, grant no. TG-AST160028, which is supported by NSF grant no. ACI-1548562.

\bibliographystyle{mnras}
\bibliography{references}


\bsp	
\label{lastpage}
\end{document}